\documentclass[11pt,letter]{article}
\usepackage{latexsym}
\usepackage{amsmath}
\usepackage{amssymb}
\usepackage{amsfonts}
\usepackage{amsthm} % clashes with llcns
\usepackage{url}
\usepackage{graphicx}
\usepackage{subfigure}
\usepackage{rotating}
\usepackage{epsfig}
\usepackage{times} % assumes new font selection scheme installed
\usepackage{setspace}
\usepackage{natbib}
\usepackage[Algorithm]{algorithm}
\usepackage[noend]{algorithmic}
\usepackage[margin=1in]{geometry}
\usepackage{tabularx}
\newcolumntype{R}{>{\raggedleft\arraybackslash}X}

\usepackage{xr}

\bibpunct{[}{]}{;}{a}{}{}

\newtheorem{Thm}{Theorem}

\newtheorem{Prob}[Thm]{Problem}

\newcommand{\ignore}[1]{}

\renewcommand{\Prob}{{\bf P}}

\newcommand{\N}{{\mathbb N}}

\newcommand{\cA}{{\mathcal A}}

\newcommand{\cC}{{\mathcal C}}

\newcommand{\cN}{{\mathcal N}}

\newcommand{\cP}{{\mathcal P}}

\newcommand{\cR}{{\mathcal R}}

\begin{document}

\title{CLEVER: Clique-Enumerating Variant Finder}
\author{Tobias Marschall$^{1*}$, Ivan Costa$^{2*}$, 
Stefan Canzar$^{1}$, 
Markus Bauer$^{3}$,\\ 
Gunnar Klau$^{1}$,
Alexander Schliep$^{5}$, 
Alexander Sch\"onhuth$^{1\dagger}$\\[2ex]
$^1$ Centrum Wiskunde \& Informatica, Amsterdam, Netherlands\\
$^2$ Federal University of Pernambuco, Recife, Brazil\\
$^3$ Illumina Inc., Cambridge, UK\\
$^4$ Rutgers, The State University of New Jersey, Piscataway, NJ, USA\\
$^*$ Joint first authorship\\
$^{\dagger}$ Corresponding author\\[2ex]
{\tt alexander.schoenhuth@cwi.nl}}

\maketitle
\thispagestyle{empty}

\begin{abstract}
  Next-generation sequencing techniques have
  facilitated a large scale analysis of human genetic variation.
  Despite the advances in sequencing speeds,
  the computational discovery of structural variants is
  not yet standard. It is likely that many
  variants have remained undiscovered in most sequenced
  individuals. 

  Here we present a novel internal segment size based approach, which
  organizes {\em all}, including also concordant reads into a {\em
    read alignment graph} where max-cliques represent maximal
  contradiction-free groups of alignments. A specifically engineered
  algorithm then enumerates all max-cliques and statistically
  evaluates them for their potential to reflect insertions or
  deletions (indels). For the first time in the literature, we compare
  a large range of state-of-the-art approaches using simulated
  Illumina reads from a fully annotated genome and present various
  relevant performance statistics.  We achieve superior performance
  rates in particular on indels of sizes 20--100, which have been
  exposed as a current major challenge in the SV discovery literature
  and where prior insert size based approaches have limitations. In
  that size range, we outperform even split read aligners. We achieve
  good results also on real data where we make a substantial amount of
  correct predictions as the only tool, which complement the
  predictions of split-read aligners.

CLEVER is open source (GPL) and available from
  \url{http://clever-sv.googlecode.com}.


\end{abstract}
\addtocounter{page}{-1}
\newpage

%======================= INTRODUCTION ==========================================
\section{Introduction}
\label{sec.intro}
%===============================================================================

\citet{HapMap2005} and \citet{1000genomes2010} have, through globally concerted
efforts, provided the first systematic view into the gamut and
prevalence of human genetic variation, including also larger genomic
rearrangements.  A staggering 8\% of the general human population have
copy number variants (CNV) affecting regions larger than
500kbp~\citep{Itsara2009}. The technology enabling this advance was
next-generation sequencing and the reduction in costs and increases of
sequencing speeds it brought along \citep{Bentley2008,Solid,Eid2009}.
The analysis of structural variation however has not kept up with the
advances in sequencing insofar as genotyping of human structural
variation has not yet become a routine procedure \citep{Alkan2011}.
Indeed it is likely that existing data sets contain structural
variations indiscoverable by current methods. These limitations are
likewise an obstacle to personalized genomics.

Here, we target \emph{deletions or insertions (indels)} between 20 and 50\,000 base pairs (bp).
In particular the discovery of indels smaller than 500\,bp is still 
challenging~\citep{Alkan2011,Mills2011a}, even in non-repetitive areas of
the genome.  That the majority of structural variants resides in
repetitive areas complicates the problem further due to the
resulting read mapping ambiguities.

\paragraph{Categorization of our and prior work.} 
A {\em (paired-end) read} is a fragment of DNA both
ends of which have been sequenced. We refer to the sequenced ends of
the read as {\em (read) ends} and to the unsequenced part of the fragment
between the two ends as {\em internal segment} or {\em insert}. An
{\em alignment} $A$ of a paired-end read is a pair of alignments of
both ends. We say that a read has been {\em multiply mapped} if it
aligns at several locations in the reference genome and {\em uniquely
  mapped} in case of only one alignment.  Existing approaches for
structural variant discovery can be classified into three broad classes:
first, those based on the read
alignment coverage, that is, the number of read ends mapping to a location
\citep{Campbell2008,Chiang2009,Alkan2009,Sudmant2010,Yoon2009,Abyzov2011},
second, those analyzing the paired-end read internal segment size
\citep{Korbel2009,Hormozdiari2009,Chen2009,Lee2009,Sindi2009,Quinlan2010}, and 
third, split-read alignments \citep{Mills2006,Ye2009}.
Refer to \cite{Medvedev2009} as well as to \cite{Alkan2011} for
reviews. 
A major difference
is that the first two classes align short reads by standard read
mappers, such as BWA \citep{Li2009a}, Mr and MrsFast
\citep{Alkan2009,Hach2010} and Bowtie \citep{Langmead2009}.
% , to only name a few most popular and recent ones, see \citep{Li2010c} for a review.
Split-read aligners however compute custom alignments which span
breakpoints of putative insertions and deletions. They usually have
advantages over insert size based approaches on smaller indels while
performing worse in predicting larger indels.

\begin{figure*}%%
\includegraphics[width=\textwidth]{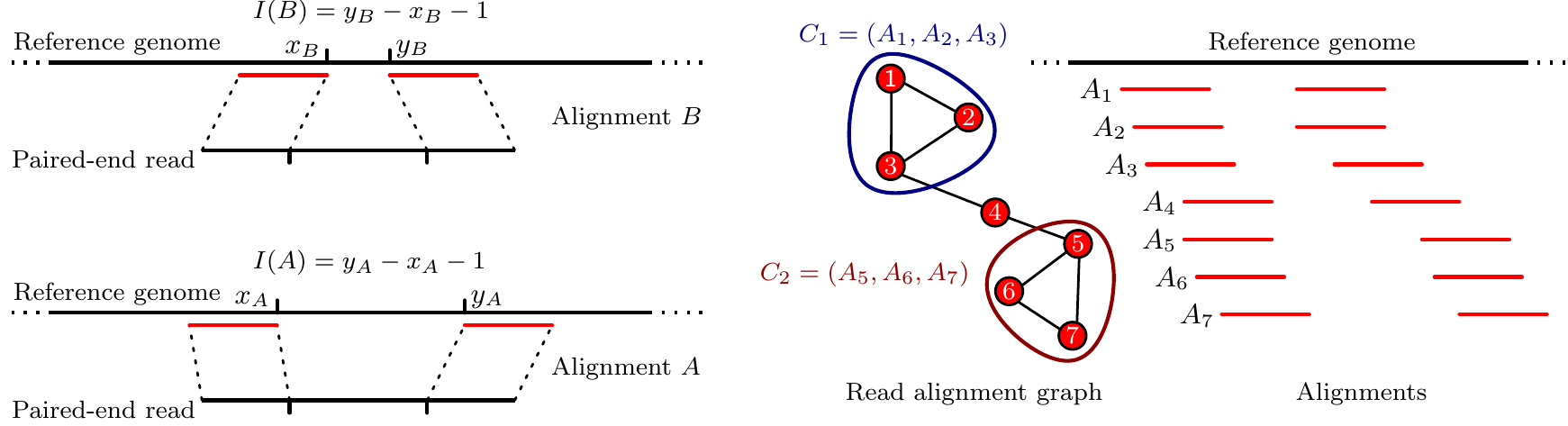}%%
  \caption{Left part: two read alignments. Assuming $I(A)>\mu>I(B)$
    where $\mu$ is the mean of the true insert size distribution,
    alignment $A$ is likely to indicate a deletion while alignment $B$ may
    indicate an insertion.  Right part: Read alignment graph for seven
    closely located read alignments. Note that
    $1/3(I(A_5)+I(A_6)+I(A_7))>1/3(I(A_1)+I(A_2)+I(A_3)$.  Assuming
    that all alignments have equal weight, $C_2$ is more likely to
    indicate a deletion than $C_1$ through a hypothesis test as in 
    equations~\eqref{eq.h0} and~\eqref{eq.h0AJ}.  Note that we have not marked
    cliques $(A_3,A_4)$ and $(A_4,A_5)$. See Fig.~\ref{fig.edges}
    for definition of edges.}
  \label{fig.aligncoord}
\end{figure*}

It is common to many library protocols that internal segment size
follows a normal distribution with machine- and protocol-specific mean
$\mu$ and standard deviation $\sigma$. On a side remark we would like
to point out that our approach does not depend on this assumption and
that we also offer an interface for arbitrary internal segment size
distributions (which may result from preparing libraries without a
size selection step, as one example) to the user.  One commonly
defines {\em concordant and discordant alignments}: an alignment with
interval length $I(A)$ (see Figure \ref{fig.aligncoord}) is concordant
iff $|I(A)-\mu| \le K\sigma$ and discordant otherwise.  The constant
$K$ can vary among the different approaches.  A {\em concordant read}
is defined to concordantly align with the reference genome, that is,
it should give rise to at least one concordant alignment.

With only one exception \citep[MoDIL]{Lee2009}, all prior approaches
discard concordant reads. In this paper, we present CLEVER, a novel
insert size based approach that takes {\em all, also concordant reads}
into consideration. While a single discordant read is significantly
likely to testify the existence of a structural variant, a single
concordant read only conveys a weak variant signals if any. Ensembles of
{\em consistent concordant alignments} however can provide significant
evidence of usually smaller variants. The major motivation of this
study is to systematically take advantage of such groups of alignments in order to not miss any
significant variant signal among concordant reads. 
% We do this to
% increase recall in particular for smaller indels and detect variants
% that cannot be discovered by existing tools.

We employ a statistical framework, which addresses deviations in
insert size, alignment quality, multiply mapped reads and coverage
fluctuations in a principled manner. As a result, our approach
outperforms all prior insert size approaches on both simulated and
real data and also compares favorably with two state-of-the-art
split-read aligners. Beyond its favorable results, our tool predicts a
substantial amount of correct indels as the only tool (for example,
more than 20\% of true deletions of $20-49$ bp in the simulated
data). Overall, CLEVER's correct calls beneficially complement those
of the split-read aligner considered \citep[PINDEL]{Ye2009}.

Moreover, we need approximately $8$ hours on a single CPU for a 30x
coverage whole-genome dataset with approximately $1$ billion reads,
which compares favorably with estimated 7,000 CPU hours needed by
MoDIL, the only method that also takes all reads into consideration.

%======================= APPROACH ==============================================
\subsection{Approach and Related Work}
\label{ssec.approach}
%===============================================================================

\subsubsection{Graph-Based Framework}

Our approach is based on organizing all read alignments into a read
alignment graph whose nodes are the alignments and edges reflect that
the reads behind two overlapping alignments are, in rigorous
statistical terms, likely to stem from the same allele.  Accordingly,
maximal cliques (max-cliques) reflect maximal consistent groups of
alignments that are likely to stem from the same location in a donor
allele. Since we do not discard alignments, the number of nodes in our
read alignment graph is large: more than $10^9$ nodes in the instances
considered here. We determine all max-cliques in this graph by means of a
specifically engineered, fast algorithmic procedure.

The idea to group alignments into location-specific, consistent
ensembles, such as max-cliques here, is not new. In fact, it has been
employed in the vast majority of previous insert size based
approaches. We briefly discuss related concepts of the three most
closely related approaches, by \citet[VariationHunter
  (VH)]{Hormozdiari2009}, \citet[GASV]{Sindi2009} and
\citet[HYDRA]{Quinlan2010}.  Although not framing it in rigorous
statistical terms, HYDRA is based on precisely the same concept of a
max-clique as our approach. After constructing the read alignment
graph from discordant reads alone, they employ a heuristic algorithm
to find max-cliques. Since no theoretical guarantee is given, it
remains unclear whether HYDRA enumerates them all. The definition of a
`valid cluster' in VH \citep{Hormozdiari2009} relaxes our definition
of a clique in a subtle, but decisive aspect.  As a consequence, each
of our max-cliques forms a valid cluster, but the opposite is not
necessarily true. The reduction in assumptions however allows VH to
compute valid clusters as max-cliques in interval graphs, in a nested
fashion, which yields a polynomial-runtime
algorithm. \citet[GASV]{Sindi2009} use a geometrically motivated
definition which allows application of an efficient plane-sweep style
algorithm.  A closer look reveals that each geometric arrangement of
alignments inferred by GASV constitutes a max-clique in our sense, but
not necessarily vice versa, even if a max-clique is formed by only
discordant read alignments. We recall that GASV, HYDRA and VH do not
consider concordant read data hence consider read alignment graphs of
much reduced sizes.

Finding maximal cliques is $\cN\cP$-hard in general graphs.
Based on the idea that the read alignment graph we consider still
largely resembles an interval graph, we provide a specifically engineered routine
that computes and tests all max-cliques in reasonable time---about 1h on a current 8~core machine for a whole human genome sequenced to 30x coverage---despite that we do not discard any read. 

\subsubsection{Significance Evaluation}\label{sec:intro_significance_evaluation}
\paragraph{Commonly Concordant and Discordant Reads.}
Testing whether $|I(A)-\mu| \le K\cdot\sigma$, to determine whether
a single alignment is concordant, is equivalent to performing a Z-test
at significance level $p_K:=1-\Phi(K)$ where $\Phi$ is the standard
normal distribution function. However, when determining whether $m$
consistent alignments (such as a clique of size $m$) with mean
interval length $\bar{I}$ are {\em commonly concordant}, a Z-test for
a sample of size $m$ is required, which translates to
\begin{equation}
  \label{eq.coconcord}
  1-\Phi(\sqrt{m}\cdot \frac{|\bar{I} - \mu|}{\sigma})\ge p_K\;\Leftrightarrow\;\sqrt{m}\cdot |\bar{I}-\mu| \le K\cdot\sigma.
\end{equation}
Due to the factor $\sqrt{m}$, already smaller deviations
$|\bar{I}-\mu|$ turn out to render the alignments {\em commonly
  discordant}. In our approach, we rigorously expand on this idea---in
a rough description, each max-clique undergoes a
Inequality-\eqref{eq.coconcord}-like hypothesis test.

\paragraph{Multiply Mapped Reads.}
While we approach the
idea of not ``overusing'' multiply mapped reads in an essentially
different fashion, our routine serves analogous purposes as the
set-cover routines of VH and HYDRA. The difference is that we
statistically control read mapping ambiguity, but do not aim at resolving it.

Following \cite{Li2008}, we compute each alignment's probability of being correctly placed.
In case of a max-clique consisting of
alignments $A_1,...,A_n$ (all from different reads) with probabilities
$p_1,...,p_n$, let $A_J,J\subset\{1,...,n\}$ be the event that
precisely the alignments $A_j,j\in J$ are correct. We compute
$\Prob(A_J)=\prod_{j\in J}p_j\prod_{j\not\in J}(1-p_j)$.  Let $H_0$ be
the null hypothesis of that the allele in question---we recall that
max-cliques just represent groups of alignments likely to be from the
same allele---coincides with the reference genome. In correspondence
to Inequality~\eqref{eq.coconcord}, we compute
\begin{equation}
  \label{eq.h0AJ}
  \Prob_{H_o}(A_J) := 1 - \Phi(\sqrt{|J|}\frac{|\bar{I}_J-\mu|}{\sigma})
\end{equation}
with $\bar{I}_J=\frac{1}{\sum_{j\in J}p_j}\sum_{j\in J}p_jI(A_j)$,
which is the probability of observing $A_j, j\in J$ when assuming the
null hypothesis, given $A_J$.  We further compute
\begin{equation}
  \label{eq.h0}
  \Prob_{H_0}(A_1,...,A_n) = \sum_{J\subset \{1,...,n\}}\Prob(A_J)\Prob_{H_0}(A_J)
\end{equation}
\begin{sloppypar}
as the probability
% [note that $\sum_{J\subset\{1,...,n\}}P(A_J)=1$]
that max-clique $A_1,...,A_m$ does {\em not} support an indel
variant. We further correct $\Prob_{H_0}(A_1,...,A_n)$ with a {\em
  local Bonferroni factor} to adjust for coverage-mediated
fluctuations in the number of implicitly performed tests. If the
corrected $\Prob_{H_0}(A_1,...,A_n)$ is significantly small, it is
likely that (at least) one allele in the donor is affected by an indel
at that location. See Methods for details. In a last step, we
apply the Benjamini-Hochberg procedure to correct for multiple
hypothesis testing overall. Note that among the prior approaches only
MoDIL \citep{Lee2009} addresses to correct for multiple hypothesis
testing (also using Benjamini-Hochberg), although many others either
explicitly (e.g.~\cite{Chen2009a}) or implicitly
(e.g.~\cite{Hormozdiari2009,Korbel2009,Quinlan2010}) perform multiple
hypothesis tests.
\end{sloppypar}

Among the statistically motivated approaches, \citet{Lee2009}, after
clustering, use Kolmogorov-Smirnov tests in combination with
bimodality assumptions, \citet{Chen2009a} measure both deviations from
Poisson-distribution based assumptions (BreakdancerMax) and use
Kolmogorov-Smirnov (BreakdancerMin) tests to discover copy number
changes.

%======================= METHODS ===============================================
\section{Methods}
\label{sec.methods}
%===============================================================================
\subsection{Notations, Definitions and Background}
\label{ssec.notations}

\paragraph{Reads and Read Alignments.} 
Let $\cR$ be a set of paired-end reads, stemming from a {\em donor
  (genome)} which have been aligned against the {\em reference
  (genome)}.  We write $A$ for a paired-end alignment, that is a pair
of alignments of the two ends of a read (see
Fig.~\ref{fig.aligncoord}) and $\cA(R)$ for the set of correctly
oriented alignments which belong to read $R$. We neglect incorrectly
oriented alignments and write $\cA=\cup_R\cA(R)$ for the set of all
alignments we consider. 
% We write $|.|$ to denote the cardinality of a set.
We assume that $|\cA(R)|\ge 1$; that is, each read we consider
gives rise to at least one well-oriented alignment.
We do not discard any reads.

We write $x_A$ for the rightmost position of the left end and $y_A$
for the leftmost position of the right end. We write $[x_A+1,y_A-1]$
and call this the {\em interval} of alignment $A$ (in slight abuse of
notation: intervals here only contains integers) and $I(A):=y_A-x_A-1$
for the {\em (alignment) interval length}. When referring to alignment
intervals, we sometimes call $x_A,y_A$ the left and right {\em
  endpoint}. See Figure \ref{fig.aligncoord} for illustrations.

\paragraph{Internal Segment Size Statistics.}
We write $I(R)$ for the true (but unknown) internal segment size of
paired-end read $R$ which is the (unknown) length of the entire read
$R$ minus the (known) lengths of its two sequenced ends. In the
datasets treated here, $I(R)$ can be assumed normally distributed with
a given mean $\mu$ and standard deviation $\sigma$
\citep{Li2008,Li2009a,Hormozdiari2009,Lee2009}, that is,
$I(R)\sim\cN_{(\mu,\sigma)}$.  Estimation of mean $\mu$ and standard
deviation $\sigma$ poses the challenge that the empirical statistics
on alignment length, further denoted as $\Prob_{\rm Emp}$ are
fat-tailed, due to already reflecting structural variation between
donor and reference. Here, we rely on robust estimation routines, as
implemented by BWA \citep{Li2009a}. Note that in general we allow to
deal with arbitrary internal segment size statistics.

\paragraph{Alignment Scores and Probabilities.}
As described by \citet{Li2008}, we determine 
$\log_{10}\Prob_{\rm Ph}(A) := -\sum_jQ_j/10$ where $j$ runs over all
mismatches in both read ends and $Q_j$ is the Phred score for position
$j$, that is $10^{-(Q_j/10)}$ is the probability that the nucleotide
at position $j$ reflects a sequencing error. Hence $\Prob_{\rm Ph}(A)$
is the probability that the substitutions in alignment $A$ are due to
sequencing errors.  The greater $\Prob_{\rm Ph}(A)$ the more likely
that $A$ is correct so $\Prob_{\rm Ph}(A)$ serves as a statistical
quality assessment of $A$. Note that to neglect SNP rates and indels
reflects common practice \citep{Li2008,Li2009a}, which is justified by
that in Illumina reads substitution error rates are higher than SNP rates, indel
sequencing error rates and DIP (deletion/insertion polymorphism) rates by orders
of magnitude \citep{Bravo2010,Albers2011}.

Patterned according to \citet{Li2008,Li2009a}, we integrate the
empirical interval length distribution $\Prob_{\rm Emp}(I(A))$ into an
overall score
$S_0(A) := \Prob_{\rm Ph}(A)\cdot \Prob_{\rm Emp}(I(A))$ and obtain as 
the probability that $A$ is the correct alignment for its read, by
application of Bayes' formula
\begin{equation}
  \label{eq.alprior}
  \Prob_0(A) = \frac{S_0(A)}{\sum_{\tilde{A}\in\cA(R)}S_0(\tilde{A})}.
\end{equation}

\paragraph{The Read Alignment Graph.}

We arrange all scored read alignments $\cA$ in form of an undirected,
weighted graph $G=(\cA,E,w)$. Since we identify nodes with read
alignments from $\cA$, we use these terms interchangeably. We draw
an edge between alignments $A,B\in\cA$ if we cannot reject the hypothesis
that, in case they are both correct, their reads can stem from the
same allele.  See the subsequent paragraph for details. The weight
function $w:\cA\to[0,1]$ is defined by $w(A) := \Prob_0(A)$.
We
further label nodes by $r:\cA\to \{1,...,N\}$ where $r(A)=n$ iff
$A\in\cA(R_n)$ that is alignment $A$ is due to read $R_n$. 

As usual, we write $\delta(A):=|\{B\in \cA\mid (A,B) \in E\}|$ for the
{\em degree} of node $A$. A {\em clique} $\cC\subset\cA$ is defined as
a subset of mutually connected nodes, that is, 
$(A,B)\in E$ for all $A,B\in \cC$. A {\em maximal clique} $\cC$ is a clique such that for
every node $A\in\cA\setminus \cC$ there is $B\in \cC: (A,B)\not\in E$.
Note that by our definition of edges, a clique is a group of
alignments which can be jointly assumed to be associated with the same
allele, or, in other words, to jointly support the same local
phenomenon in the donor genome. Maximal cliques are obviously
particularly interesting: while all alignments in the clique are
likely to support the same local phenomenon, joining any other {\em
  overlapping} alignment may lead to conflicts. 
% Hence a maximal
% clique is likely to commonly support a certain
% phenomenon in the donor genome.

\paragraph{Edge Computation.}

\begin{figure*}%%
  \includegraphics[width=\textwidth]{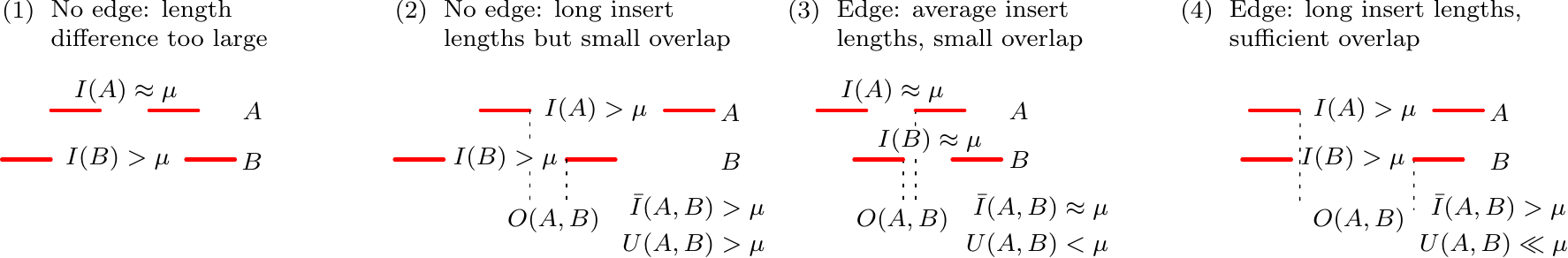}
  \caption{Four scenarios of two overlapping alignment pairs $A$ and $B$.  In the \emph{read alignment graph}, two alignments
    are connected by an \emph{edge} if they are compatible, that is,
    they support the same phenomenon.  (1)~Alignment~$A$ has an insert
    length about the expected insert length $\mu$, suggesting that there
    is no variation present but alignment~$B$ has an insert length much
    larger than $\mu$ suggesting a deletion. Hence, $A$ and $B$ are not
    compatible. (2)~Both alignments have similar insert lengths larger
    than $\mu$, both suggesting a deletion of size $I(A)-\mu\approx
    I(B)-\mu$, but the overlap $O(A,B)$ is too small to harbor a
    deletion of this size. Thus, they are incompatible. (3)~Both
    alignments do not suggest any variation and are therefore
    compatible. (4)~Similar to Case~(2), but now the overlap is large
    enough to contain the putative deletion.} 
  \label{fig.edges}
\end{figure*}

See Figure \ref{fig.edges} for illustrations of the following.  Let
$A,B$ be two alignments. 
We define:

{\bf -} $\Delta(A,B):= |I(A)-I(B)|$ is the absolute difference of
interval length.

{\bf -} $O(A,B):=\min(y_A,y_B) - \max(x_A,x_B)-1$ where in case of
$O(A,B)\ge 0$ we refer to all positions between $\max(x_A,x_B)$ and
$\min(y_A,y_B)$ as their {\em common interval}. 
% Note that $O(A,B)<0$
% reflects that $A$ and $B$'s internal segments cannot commonly harbor
% a variant.

{\bf -} $\bar{I}(A,B):=(I(A)+I(B))/2$ is the {\em mean interval lengths}.

{\bf -} $U(A,B):= \bar{I}(A,B) - O(A,B)$ is the difference of mean
interval length and overlap. To motivate this quantity, note that, in
case $A$ and $B$ overlap (hence length of common interval $O(A,B)>0$)
and are from the same allele, a deletion at that location can only
happen to take place in their common interval. If $U(A,B)$ is large
then $\bar{I}(A,B)$ significantly deviates from $\mu$ and the common
interval is not large enough to explain this by a large enough
deletion. Hence it is unlikely $A,B$ are from the same allele.

Let $X$ be $\cN_{(0,1)}$-distributed and, as above, $\mu, \sigma$ be
mean and variance of the insert size distribution. We draw an edge
between alignments $A,B$ in the read alignment graph iff the reads
of $A$ and $B$ are different, $O(A,B)\ge 0$
and
\begin{align}
  \label{eq.edgetest}
  \Prob&(|X|\ge \frac{1}{\sqrt{2}}\frac{\Delta(A,B)}{\sigma}) \le 0.05\quad\text{and}\\
  \label{eq.edgetest2}
  \Prob&(X\ge \sqrt{2}\frac{(U(A,B)-\mu)}{\sigma}) \le 0.05.
\end{align}
%for $p_1,p_2$ being appropriately chosen p-values.
Inequality~\eqref{eq.edgetest} is a two-sided two sample Z-test to measure {\em
  statistically compatible insert size}.  Inequality~\eqref{eq.edgetest2}
reflects a one-sided one-sample Z-test for {\em statistically
  consistent overlap} \citep{Wasserman}.  If two alignments $A,B$ with
$O(A,B)\ge 0$ pass these tests, we have no reason to reject the
hypothesis that the alignments are from the same allele so we draw an
edge.  
% We recall that similar ideas have been presented by
% \citet[HYDRA]{Quinlan2010}.

\subsection{CLEVER: Algorithmic Workflow}
\label{ssec.clever}
$\/$\indent{\bf 1.} Enumerating Maximal Cliques: We compute all {\em maximal cliques} in
the read alignment graph.

{\bf 2.} We assign two p-values $p_D(\cC),p_I(\cC)$ to each maximal
clique $\cC$ which are the probabilities that the alignments
participating in $\cC$ do not commonly support a deletion or
insertion. So the lower $p_D(\cC)$ or~$p_I(\cC)$, the more likely it is
that $\cC$ supports a deletion or insertion, respectively.
 
{\bf 3.} For the thus computed p-value, we control the false discovery
rate at 10\,\% by applying the standard Benjamini-Hochberg procedure
separately for insertions and deletions.  All cliques remaining after
this step are deemed \emph{significant} and processed further.

{\bf 4.} Determining Parameters: We parametrize deletions $D$ by their
left breakpoint $D_B$ and their length $D_L$, which denotes that
reference nucleotides of positions $D_B,...,D_B+D_L-1$ are missing in
the donor.  We parametrize insertions $I$ by their breakpoint $I_B$
and their length $I_L$ such that before position $I_B$ in the
reference there has been a sequence of length $L$ inserted in the
donor.  Depending on whether $\cC$ represents a deletion or insertion,
we determine [$w(\cC):=\sum_{A\in\cC}w(A)$]
\begin{equation}
  \frac{1}{w(\cC)}\sum_{A\in\cC}w(A)(I(A)-\mu)\quad\text{resp.}\quad\frac{1}{w(\cC)}\sum_{A\in\cC}w(A)(\mu-I(A))
\end{equation}
as the length $D_L$ of the deletion resp.~$I_L$ of the insertion.
We determine breakpoints $D_B$ or $I_B$ such that the predicted deletion
or insertion sits right in the middle of the intersection of all internal segments of alignments in $\cC$. 

\paragraph{Enumerating Maximal Cliques.}
%An \emph{interval graph} is given by a set of intervals on the real
%line.  Its maximal cliques can be found efficiently in time linear in
%the size of the graph \citep{Fishburn}.  
%This algorithm is applied by
%\cite{citeulike:7263200} to find maximal valid clusters for copy
%events \todo{What are copy events?}.  
%\cite{Hormozdiari2009} characterize valid clusters for insertion and
%deletion events by overlapping intervals of similar size. Both
%requirements can be modeled naturally by intervals that must intersect
%and thus maximal cliques can be determined again by applying the above
%mentioned algorithm twice.

We identify nodes of the read alignment graph with the intervals of the
corresponding alignments. We first sort the $2m$ endpoints of these
intervals, $m:=|\mathcal{A}|$, in ascending order of their positions. We
then scan this list from left to right. We maintain a set of
\emph{active} cliques that could potentially be extended by a
subsequent interval, which initially is empty. If the current element
$\ell$ of the list is a left endpoint, we extend the set of active
cliques according to the following rules. For the sake of simplicity,
let us assume that a unique interval starts at $\ell$, corresponding to
a vertex $A$ in the read alignment graph $G$.  Let $N(A)$ be the open
neighborhood of $A$.  If $\mathcal{C}\cap N(A)=\emptyset$ for all
active cliques $\mathcal{C}$, add a singleton clique $\{A\}$ to the
set of active cliques.  Otherwise, for each active clique
$\mathcal{C}$,
\begin{itemize}
 \item[(i)] if $\mathcal{C}\cap N(A)=\mathcal{C}$, then $\mathcal{C}:=\mathcal{C}\cup\{A\}$, otherwise
 \item[(ii)] if $\mathcal{C}\cap N(A)\neq\emptyset$, add $(\mathcal{C}\cap N(A))\cup\{A\}$ to the set of active
       cliques.
\end{itemize}
Finally, duplicates and cliques that are subsets of others
are removed. 

If the current element $\ell$ of the list is a right endpoint, we
output all cliques that contain at least one interval ending at
$\ell$.  These cliques go out of scope and are thus maximal. We remove
intervals ending at $\ell$ from active cliques. Cliques that become
empty are removed from the set of active cliques.

\paragraph{Runtime Analysis.} Let $k$ be an upper bound on local alignment
coverage, $c$ be the maximum number of active cliques and $s$ be the
size of the output. The detailed runtime analysis of
section~\ref{sec:algor-engin} in the Supplement gives a total running
time of $\mathcal{O}(m(\log m+kc^2)+s)$. Despite these rather moderate
worst case guarantees, however, our algorithm is very fast in
practice. See again the supplementary section~\ref{sec:algor-engin}
for an analysis of the corresponding reasons.

\paragraph{P-Values for Cliques.}
We proceed as sketched in the Section~\ref{sec:intro_significance_evaluation}.
Let $\cC$ be a maximal clique in the read alignment graph and let
$w(\cC) := \sum_{A\in\cC}w(A) = \sum_{A\in\cC}\Prob_0(A)$
be the {\em the weight of the clique}. Let
$\bar{I}(\cC) := \frac{1}{w(\cC)}\cdot \sum_{A\in\cC}w(A)\cdot I(A)$
be the {\em weighted mean of alignment interval length} of the clique.
Let $\Phi$ be the standard normal distribution function. Let
$\rho(\cC)$ be the number of alignments which are at the genomic
location of the clique. For example, in Figure \ref{fig.aligncoord},
$\rho(C_1)=\rho(C_2)=7$ is just the number of alignments which overlap
with one another at this position of the reference.
We compute
\begin{align}
  \label{eq.indeltest1}
  p(\cC)_D&:=2^{\rho(\cC)}\sum_{J\subset \cC}\Prob_{H_0}(A_J)[1-\Phi(\sqrt{|J|}\frac{\bar{I}(\cC)-\mu}{\sigma})]\\
  p(\cC)_I&:=2^{\rho(\cC)}\sum_{J\subset \cC}\Prob_{H_0}(A_J)[\Phi(\sqrt{|J|}\frac{\bar{I}(\cC)-\mu}{\sigma})]\label{eq.indeltest2}
\end{align}
just as in equations~\eqref{eq.h0} and~\eqref{eq.h0AJ} with the difference that we
distinguish between cliques which give rise to deletions and
insertions.  $2^{\rho(\cC)}$ is the number of subsets of alignments
one can test at this location, that is the virtual number of tests
which we perform, so multiplying by $2^{\rho(\cC)}$ is a
Bonferroni-like correction. This correction accounts for coverage
fluctuations.

 If $p(\cC)_D$ is significantly small
then $\bar{I}(\cC)$ is significantly large, hence the alignments in
$\cC$ are deemed to commonly support a deletion. Analogously, if
$p(\cC)_I$ is significantly small, then $\cC$ is supposed to support
an insertion.
Refer to Supplement~\ref{app.approx} for details on how the exponential sums in equations~\eqref{eq.indeltest1} and~\eqref{eq.indeltest2} can be computed efficiently.

%======================= RESULTS ===============================================
\section{Results and Discussion}
\label{sec.results}
%===============================================================================

  \paragraph{\bf Simulation: Craig Venter Reads.} We downloaded the comprehens\-ive
  set of annotations of both homozygous and heterozygous structural
  variants (also including inversions and all other balanced
  re-arrangements) for Craig Venter's genome, as documented by
  \citet{Levy2007} and introduced them into the reference genome, thereby
  generating two different alleles. If nested effects lead to ambiguous
  interpretations we opted for an order which respects the overall
  predicted change in copy number. We used UCSC's
  SimSeq\footnote{https://github.com/jstjohn/SimSeq} as read simulator
  to simulate Illumina paired-end reads with read end length $100$ at
  coverage $15x$ for each of the two alleles which yields $30x$ sequence
  coverage overall.  

  \paragraph{\bf Real Data: NA18507.}  We further were provided with
  reads of the genome of an individual from the Yoruba in Ibadan,
  Nigeria by Illumina. Reads were sequenced on a GAIIx and are now
  publicly
  available\footnote{ftp://ftp.sra.ebi.ac.uk/vol1/ERA015/ERA015743/srf/}. Read
  ends are of length $101$. Read coverage is $30x$. For benchmarking
  purposes, we used annotations from \cite[Gen.Res.]{Mills2011a}
  merged with ``DIP'' annotations from the HGSV
  Project\footnote{http://hgsv.washington.edu} database.

  \paragraph{\bf Reference Genome and Alignments}
  As a reference genome, we used
  version hg 18, as downloaded from the UCSC Genome
  Browser. All reads considered were
  aligned using BWA \citep{Li2009a} with the option to allow $25$
  alignments per read end, which amounts to a maximum of $25^2$
  alignments per paired-end read. BWA determined mean insert size
  $\mu\approx 112$ and standard deviation $\sigma\approx 15$ for both
  simulated and NA18507 reads. Note that re-alignment of discordant
  reads with a slow, but more precise alignment tool, such as
  Novoalign\footnote{http://www.novocraft.com/main/index.php} can lead
  to subsequent resolution of much misaligned sequence and therefore has
  been suggested by \cite{Quinlan2010}. We are
  aware that all methods considered would benefit from such
  (time-consuming) re-alignment of reads.

\begin{table*}
  \caption{\small This table shows benchmarking results for simulated
    (Venter) and real data (NA18507). Performance rates as recall,
    precision, exclusive predictions (Exc. which are true predictions,
    uniquely predicted by that tool) and F-measure are grouped by
    different indel size ranges. A dash resp.~N/A indicates no
    prediction resp.~not applicable in that category (GASV cannot
    report insertions and SV-seq can only predict insertion
    breakpoints but not their length such that one cannot evaluate
    them by common criteria). Insertions significantly exceeding the
    internal segment size ($\approx 112$ here) cannot be detected by
    insert size based approaches. PINDEL does not detect such
    insertions either.}\label{tab:results}
\vspace{-.5em}
\begin{center}\small

\begin{tabularx}{\textwidth}{lRRRRp{1pt}RRRRp{10pt}RRRRRR}
\hline
\textbf{Data Set}    & \multicolumn{4}{c}{\textbf{Venter Insertions}} & & \multicolumn{4}{c}{\textbf{Venter Deletions}} & & \multicolumn{3}{c}{\hspace{-.5em}\textbf{NA18507 Insertions}} & \multicolumn{3}{c}{\hspace{-.3em}\textbf{NA18507 Deletions}} \\
                     & Prec. & Rec. & Exc. & F & & Prec. & Rec. & Exc. & F & & RPr. & Rec. & Exc. & RPr. & Rec. & Exc.\\
\hline
\multicolumn{10}{l}{\textbf{Length Range 20--49} (8,786 true ins., 8,502 true del.)} & & \multicolumn{6}{l}{(2,295 true ins., 2,192 true del.)}\\
CLEVER          &          62.5  &  \textbf{53.0} &  \textbf{20.4} &  \textbf{57.4} & &         60.4  &  \textbf{66.8} &  \textbf{15.9} &  \textbf{63.4} & &  \textit{7.7}  & \textit{24.1}  &  \textit{8.4}  &  \textit{8.9}  & \textit{44.7}  & \textit{6.6}  \\[.3em]
BreakDancer     &             -- &           5.1  &           0.1  &             -- & &\textit{75.5}  &           7.5  &           0.0  &          13.6  & &             -- &           0.3  &           0.0  &           8.2  &           5.8  &           0.0  \\
GASV            &            N/A &           N/A  &           N/A  &            N/A & &          5.4  &          25.8  &           1.8  &           8.9  & &            N/A &           N/A  &           N/A  &           1.0  &          20.1  &           2.0  \\
HYDRA           &           0.0  &           0.0  &           0.0  &             -- & &            -- &           0.1  &           0.0  &             -- & &           0.0  &           0.0  &           0.0  &             -- &           0.0  &           0.0  \\
VariationHunter &          32.4  &           8.4  &           0.2  &          13.4  & &         66.3  &           8.0  &           0.3  &          14.3  & &           0.8  &           3.8  &           0.4  &           4.6  &           4.6  &           0.3  \\[.3em]
PINDEL${}^*$    &  \textbf{66.1} &          44.9  &          18.7  &          53.5  & &         49.5  &          55.8  &          12.1  &          52.5  & &  \textbf{13.1} &  \textbf{40.0} &  \textbf{25.3} &           9.3  &  \textbf{64.9} &  \textbf{26.3} \\
SV-seq2${}^*$   &            N/A &           N/A  &           N/A  &            N/A & & \textbf{96.0} &           1.2  &           0.0  &           2.3  & &            N/A &           N/A  &           N/A  &  \textbf{15.2} &           1.6  &           0.2  \\
\hline
\multicolumn{10}{l}{\textbf{Length Range 50--99} (2,024 true ins., 1,822 true del.)} & & \multicolumn{6}{l}{(303 true ins., 294 true del.)}\\
CLEVER          &          60.4  &  \textbf{86.6} &   \textbf{7.3} &  \textbf{71.2} & &         72.7  &  \textbf{80.7} &   \textbf{6.8} &  \textbf{76.5} & &           1.6  &  \textbf{70.3} &   \textbf{6.9} &           5.5  &  \textbf{79.6} &  \textbf{12.2} \\[.3em]
BreakDancer     &  \textbf{86.5} &          56.5  &           0.2  &          68.3  & & \textbf{87.3} &          48.1  &           0.3  &          62.0  & &  \textit{6.4}  &          15.5  &           0.0  &  \textit{9.8}  &          44.2  &           0.7  \\
GASV            &            N/A &           N/A  &           N/A  &            N/A & &         46.1  &          35.0  &           1.5  &          39.8  & &            N/A &           N/A  &           N/A  &           2.3  &          34.7  &           1.0  \\
HYDRA           &           0.0  &           0.0  &           0.0  &             -- & &            -- &           5.2  &           0.0  &             -- & &           0.0  &           0.0  &           0.0  &             -- &           2.4  &           0.0  \\
VariationHunter &          55.8  &          76.6  &           1.4  &          64.5  & &         66.5  &          65.8  &           1.5  &          66.1  & &           1.4  &          62.7  &           2.3  &           4.3  &          57.1  &           1.4  \\[.3em]
PINDEL${}^*$    &          77.5  &          20.5  &           0.3  &          32.5  & &         72.5  &          37.5  &           0.4  &          49.4  & &  \textbf{10.8} &          29.7  &           1.3  &           8.3  &          43.9  &           0.3  \\
SV-seq2${}^*$   &            N/A &           N/A  &           N/A  &            N/A & &         83.6  &          19.8  &           0.2  &          32.0  & &            N/A &           N/A  &           N/A  &   \textbf{9.9} &          28.6  &           0.3  \\
\hline
\multicolumn{10}{l}{\textbf{Length Range 100--50\,000} (3,101 true ins., 2,996 true del.)} & & \multicolumn{6}{l}{(165 true ins., 414 true del.)}\\
CLEVER          &  \textbf{66.2} &          23.8  &           2.0  &          35.1  & & \textbf{87.6} &  \textbf{69.9} &   \textbf{4.1} &  \textbf{77.7} & &           0.5  &          31.5  &           1.8  &           4.8  &  \textbf{70.3} &   \textbf{2.7} \\[.3em]
BreakDancer     &          61.0  &          17.6  &           3.0  &          27.4  & &         65.8  &          57.7  &           0.0  &          61.5  & &           0.9  &          23.0  &           1.8  &  \textit{5.2}  &          62.1  &           0.5  \\
GASV            &            N/A &           N/A  &           N/A  &            N/A & &          0.9  &          49.2  &           1.0  &           1.7  & &            N/A &           N/A  &           N/A  &           0.1  &          57.7  &           2.4  \\
HYDRA           &           0.0  &           0.0  &           0.0  &             -- & &         72.8  &          56.8  &           0.4  &          63.8  & &           0.0  &           0.0  &           0.0  &           2.0  &          65.5  &           0.5  \\
VariationHunter &          60.4  &  \textbf{25.5} &   \textbf{3.5} &  \textbf{35.8} & &         58.8  &          65.1  &           1.5  &          61.8  & &   \textbf{1.8} &  \textbf{44.9} &  \textbf{10.9} &           3.0  &          70.0  &           1.4  \\[.3em]
PINDEL${}^*$    &             -- &           1.9  &           0.0  &             -- & &         84.7  &          39.5  &           0.1  &          53.9  & &             -- &           0.6  &           0.0  &   \textbf{5.9} &          51.9  &           0.2  \\
SV-seq2${}^*$   &            N/A &           N/A  &           N/A  &            N/A & &         81.6  &          37.5  &           0.3  &          51.3  & &            N/A &           N/A  &           N/A  &           3.9  &          34.5  &           0.0  \\
\hline
\end{tabularx}
\end{center}
\vspace{-1.5em}
\end{table*}

\paragraph{\bf Experiments.} For benchmarking, we considered $5$
different state-of-the-art insert size based approaches, $4$ of which
are applicable for a whole-genome study: GASV \citep{Sindi2009},
VariationHunter \citep[v3.0]{Hormozdiari2009}, Breakdancer
\citep{Chen2009} and HYDRA \citep{Quinlan2010}. We ran MoDIL
\citep{Lee2009} only on chromosome 1 of the simulated data which, on
our machines required several hundred CPU hours. In contrast, we
process chromosome 1 in less than one hour. We also consider the
split-read aligners PINDEL \citep{Ye2009} and SV-seq2
\citep{Zhang2012}. Details on program versions and on how we ran each
method are collected in Supplement~\ref{sec:pipeline}.  In case of
deletions, we define a {\em hit} as a pair of a true deletion and a
predicted deletion which overlap and whose lengths do not differ by
more than 100bp, which roughly is the mean of internal segment
size. We say that a true insertion $(B_0,L_0)$ and a predicted
insertion $(B_1,L_1)$, where $B$ is for breakpoint, $L$ is for length,
{\em hit} each other if the intervals $[B_0+1,...,B_0+L_0]$ and
$[B_1+1,...,B_1+L_1]$ overlap. This ``overlap criterion'' precisely
parallels the one for deletions: if one views deletions in the
reference as insertions in the donor then the deletions in the
reference (relative to reference coordinates) hit {\em if and only if}
the insertions in the donor hit (relative to donor
coordinates). Again, we also require $|L_0-L_1|\le 100$. We also offer
results on alternative hit criteria which, instead of overlap, depend
on fixed thresholds on breakpoint distance and differences of indel
length in Supplement~\ref{sec:fixed-distance-stats}.  As usual, {\em
  recall} = TP/(TP+FN) where TP (= True Positives) is the number of
true deletions being hit and FN (= False Negatives) is the number of
true deletions not being hit.  For {\em Precision} = TP/(TP+FP) where
here TP is the number of predicted indels being hit and FP is the
number of predicted indels not being hit. We relate recall and
precision to one another and also display F = 2*{\em Recall}*{\em
  Precision}/({\em Recall} + {\em Precision}) [F-measure], as a common
overall statistic for performance evaluation.  We refer to {\em Exc.}
(= exclusive) as the percentage of true annotations which were {\em
  exclusively (and correctly)} predicted by the method in question.
Since the annotations for the real data set are obviously still far
from complete, a false positive may in fact point out a missing
annotation(!). We therefore call the ratio TP/(TP+FP) {\em relative
  precision (RPr.)}. For recall on the real data note that a good
amount of existing annotations may be of limited
reliability. Therefore, we refrain from displaying the on real data
meaningless F-measure rates.  Last but not least, we present average
deviation of breakpoint placement and differences in length for all
tools in the Supplement~\ref{sec:accuracy-stats}.  In
Supplement~\ref{sec:simseq}, we present CLEVER's results on simulated
data when including {\em true alignments} in the BAM files, or even
using {\em only true alignments} so as to analyze its behavior
relative to removal of external sources of errors.

\paragraph{\bf Results.}  See Table~\ref{tab:results} for performance
figures. Boldface numbers designate the best approach, italic numbers
the best insert size based approach (if not the best approach
overall). Comparing absolute numbers of true indels in the real data
with the simulated data points out immediately that the vast majority
of annotations seemingly is still missing. Therefore, all results on
the real data, in particular those on precision, can only reflect
certain trends. For the simulated data, all values reflect the
truth. As expected, performance rates greatly depend on the size of
the indels. For prediction of indels of $<20$ bp,
split-read based approaches and/or read alignment tools themselves
are the option of choice.

\paragraph{20-49 bp.} CLEVER outperforms all other approaches on the
simulated data and is the best insert size based approach also on the
real data.  PINDEL achieves best rates on the real data. Also, CLEVER
makes a substantial amount of exclusive calls in all categories. An
additional look at the tables in the Supplement, subsection
\ref{ssec:relaxed} points out that $80-90\%$ of CLEVER's indel calls
come {\em significantly} close to a real indel. Further analyses
(Supplement, section~\ref{sec:simseq}) demonstrate that $30\%$ of
CLEVER's false positives are due to misalignments and mapping
ambiguities (see also External Error Sources below). Obviously many of
those extremely close, but not truly hitting calls are due to external
errors. While Breakdancer makes little calls, it achieves high
precision which comes at the expense of reduced accuracy in
terms of indel breakpoint placement and length (see Supplement,
section~\ref{sec:accuracy-stats}).

\paragraph{50-99 bp.} Here, CLEVER achieves substantially better
recall and more exclusive calls than PINDEL also on the real data. On
the simulated data, CLEVER again achieves best overall performance.
In contrast to $20-49$ bp however, Breakdancer and VariationHunter
(VH) here already make significant contributions. While VH achieves
good overall performance, Breakdancer mostly excels in precision. As
before, when allowing a certain offset of breakpoints
(Suppl.~ssec.~\ref{ssec:relaxed}) or when integrating
correct alignments (Suppl.~sec.~\ref{sec:simseq})), CLEVER's precision
substantially rises, from $60-72\%$ to $72-96\%$ across the
categories.

\paragraph{100-50000 bp.} 
Also on indels $\ge 100$ bp CLEVER achieves most favorable performance
rates while also other tools (Breakdancer, Hydra, VariationHunter)
make decisive contributions. This documents that the current
challenges for indel discovery are rather have been the size range of
$20-100$ bp.

\paragraph{MoDIL.} We compared MoDIL with all other tools on
chromosome 1 alone, because of the excessive runtime requirements of
MoDIL. See Supplement, section~\ref{sec:modil}. Overall, MoDIL incurs
certain losses in performance with respect to CLEVER, across all
categories, but outperforms the other insert size based approaches
apart from larger indels ($\ge 100$ bp). It is noteworthy that MoDIL
makes a substantial amount of exclusive calls for insertions of
$50-99$ bp. In terms of runtime, CLEVER outperforms MoDIL by a factor
of $\approx 1000$.

\paragraph{Accuracy of Breakpoint and Length Predictions.} 
See section~\ref{sec:accuracy-stats} for related numbers. The
split-read based approaches outperform the insert size based
approaches. Among the insert size based approaches, CLEVER and GASV
are most precise for $20-49$ and $100-50000$ bp. For $50-99$ bp calls,
Breakdancer achieves favorable values.

\paragraph{External Sources of Errors.} 
See Supplement, section~\ref{sec:simseq} for related results and a
detailed discussion on to what degree misalignments and multiply
mapped reads/alignment hamper computational SV discovery.
% We 
%tested CLEVER on a dataset consisting of only the correct alignments
%of the simulated reads as well as on adding the correct alignments to
%the BWA alignments in case they had been missed by BWA
%($?\%$). Results point out that CLEVER's performance significantly
%rises when removing read mapping and Phred score quality issues. In
%particular for indels of size $\ge 50$ bp and also for smaller
%insertions, CLEVER achieves supreme performance rates (precision
%$\approx 95\%$). This points out that read mapping and machine
%parameter issues still significantly hamper computational SV
%discovery.

\paragraph{Conclusion.} 
We have presented a novel internal segment size based approach for
discovering indel variation from paired-end read data.  In contrast to
all previous, whole-genome-applicable approaches, our tool takes all
concordant read data into account.  We outperform all prior insert
size based approaches on indels of sizes $20-99$ bp and we also
achieve favorable values for long indels. We outperform the split-read
based approaches considered on medium-sized ($50-99$ bp) and larger
($\ge 100$ bp) indels. In addition, our approach detects a substantial
amount of variants missed by all other approaches, in particular in
the smallest size range considered ($20-49$ bp). In conclusion, CLEVER
makes substantial contributions to SV discovery in particular in the
size range $20-99$ bp.

Our approach builds on two key elements: first, an algorithm that
enumerates maximal, statistically contradiction-free ensembles as
max-cliques in read alignment graphs in short time and, second, a
sound statistical procedure that reliably calls max-cliques which
indicate variants.  Our approach is generic with respect to choices of
variants; max cliques in the read alignment graphs can also reflect
other variants such as inversions or translocations. As future work,
we are planning to predict inversions and to incorporate split read
information in a unifying approach.

\appendix

\section{Appendix: Engineering the CLEVER algorithm}

\label{sec:algor-engin}

%\subsection{Histograms for data set Venter}
\begin{figure}[h!]
\begin{center}
\includegraphics[width=.7\textwidth]{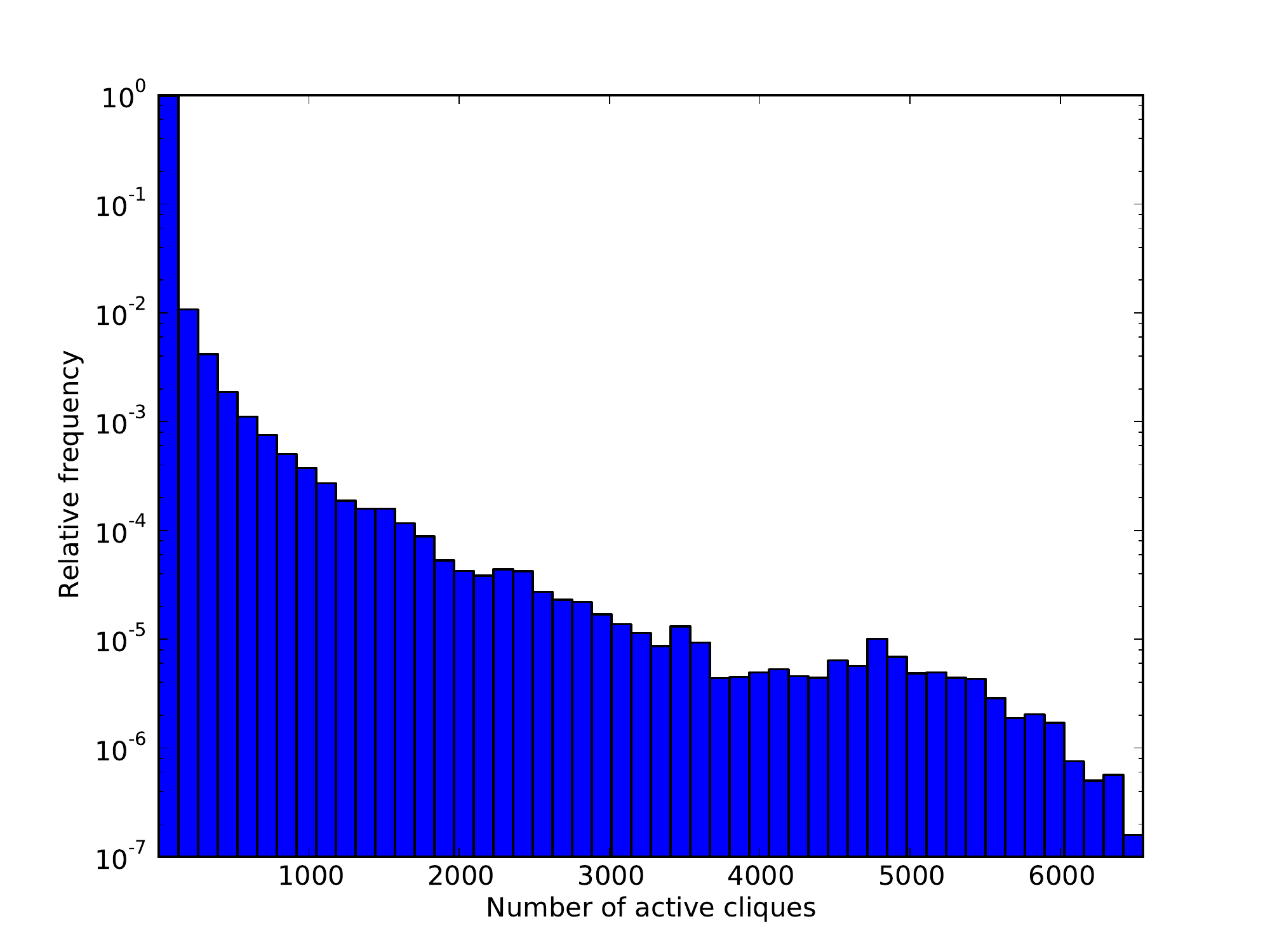}
\end{center}
\caption{Histogram of active cliques for data set Venter.}\label{fig:venter_active_cliques}
\end{figure}

\begin{figure}[h!]
\begin{center}
\includegraphics[width=.7\textwidth]{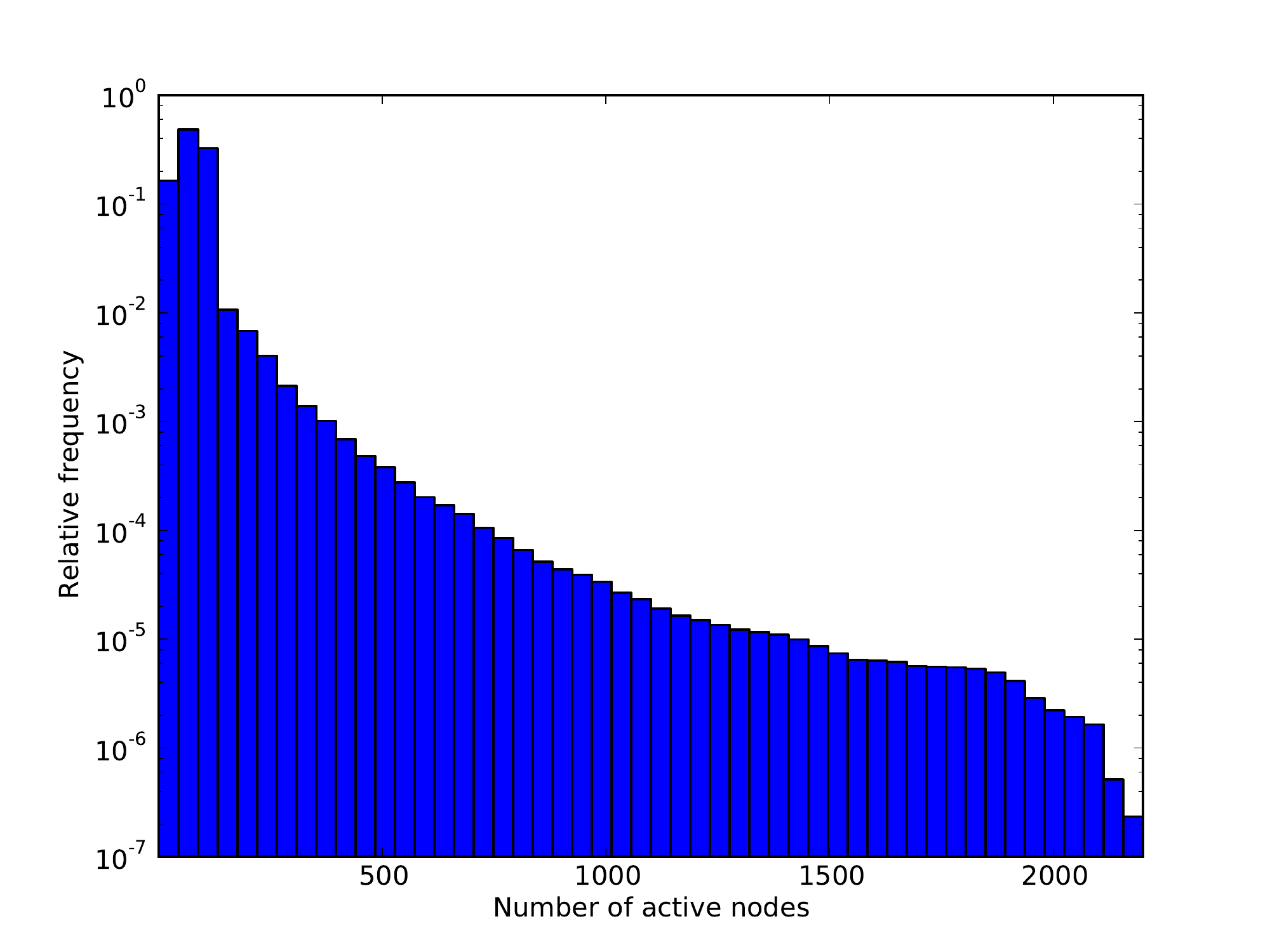}
\end{center}
\caption{Histogram of active nodes for data set Venter.}\label{fig:venter_active_nodes}
\end{figure}

\begin{figure}[h!]
\begin{center}
\includegraphics[width=.7\textwidth]{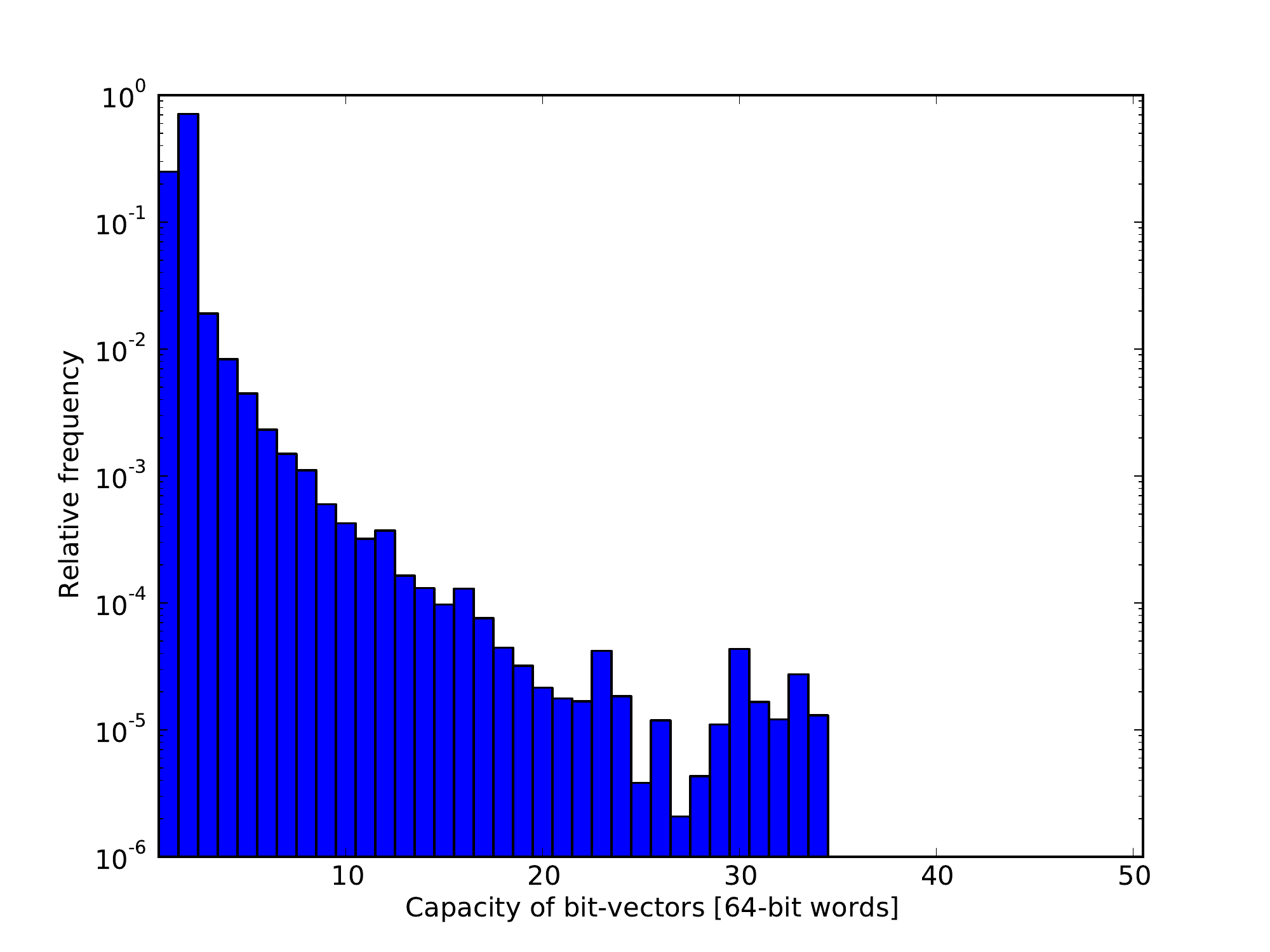}
\end{center}
\caption{Histogram of bit-vector capacities for data set Venter.}\label{fig:venter_bv_capacity}
\end{figure}

%\subsection{Histograms for data set NA18507}
\begin{figure}[h!]
\begin{center}
\includegraphics[width=.7\textwidth]{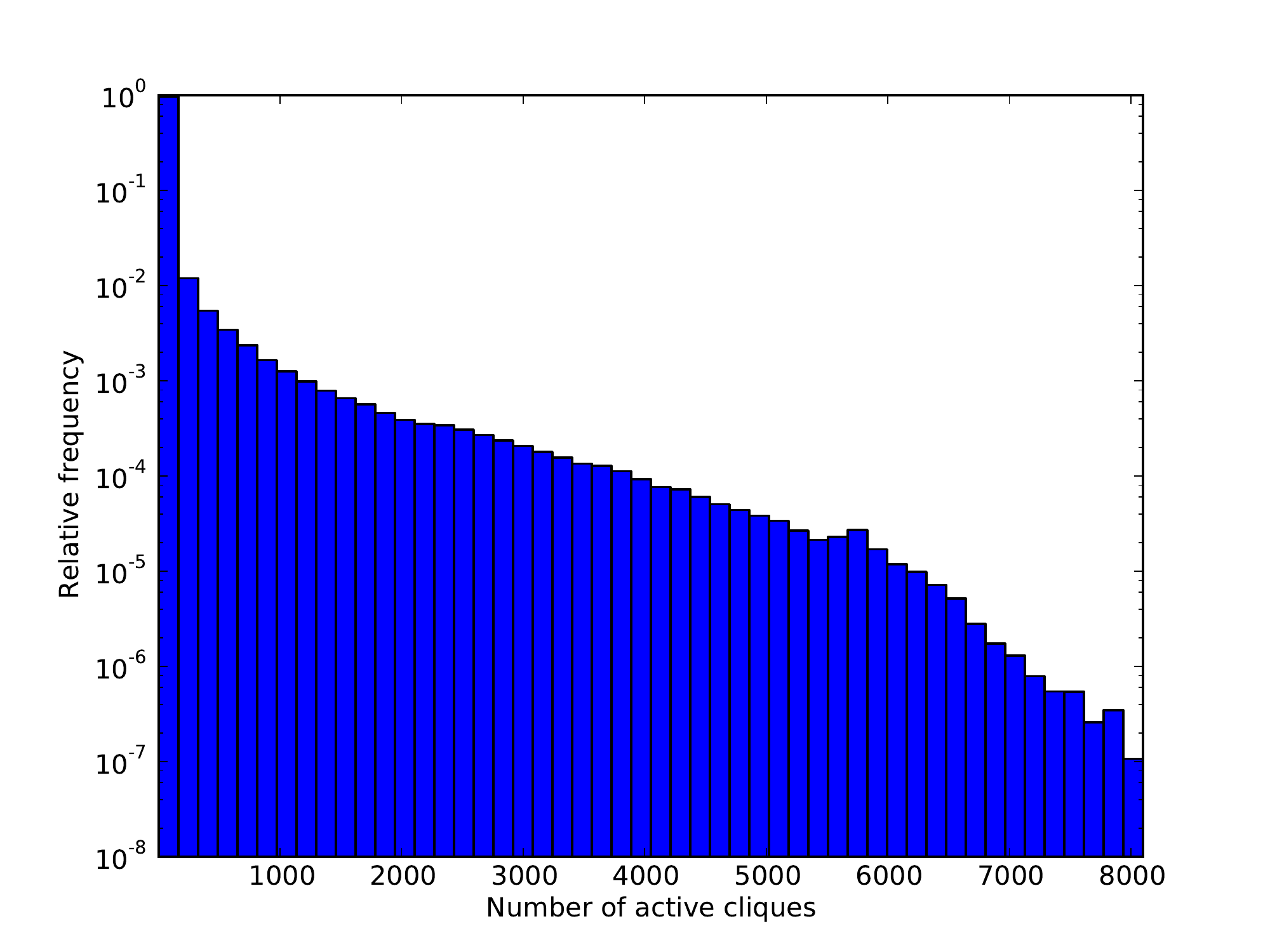}
\end{center}
\caption{Histogram of active cliques for data set NA18507.}\label{fig:NA18507_active_cliques}
\end{figure}

\begin{figure}[h!]
\begin{center}
\includegraphics[width=.7\textwidth]{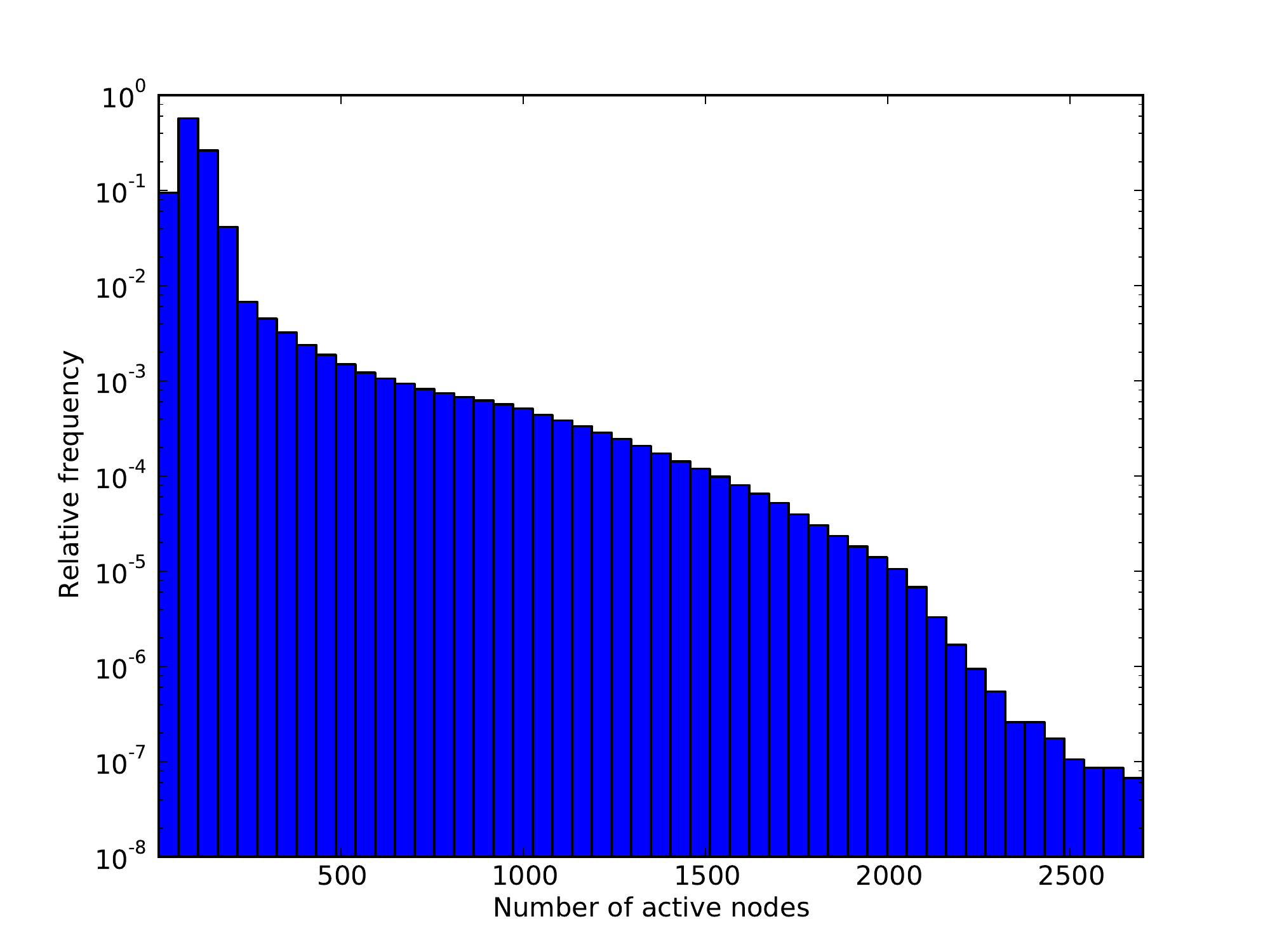}
\end{center}
\caption{Histogram of active nodes for data set NA18507.}\label{fig:NA18507_active_nodes}
\end{figure}

\begin{figure}[h!]
\begin{center}
\includegraphics[width=.7\textwidth]{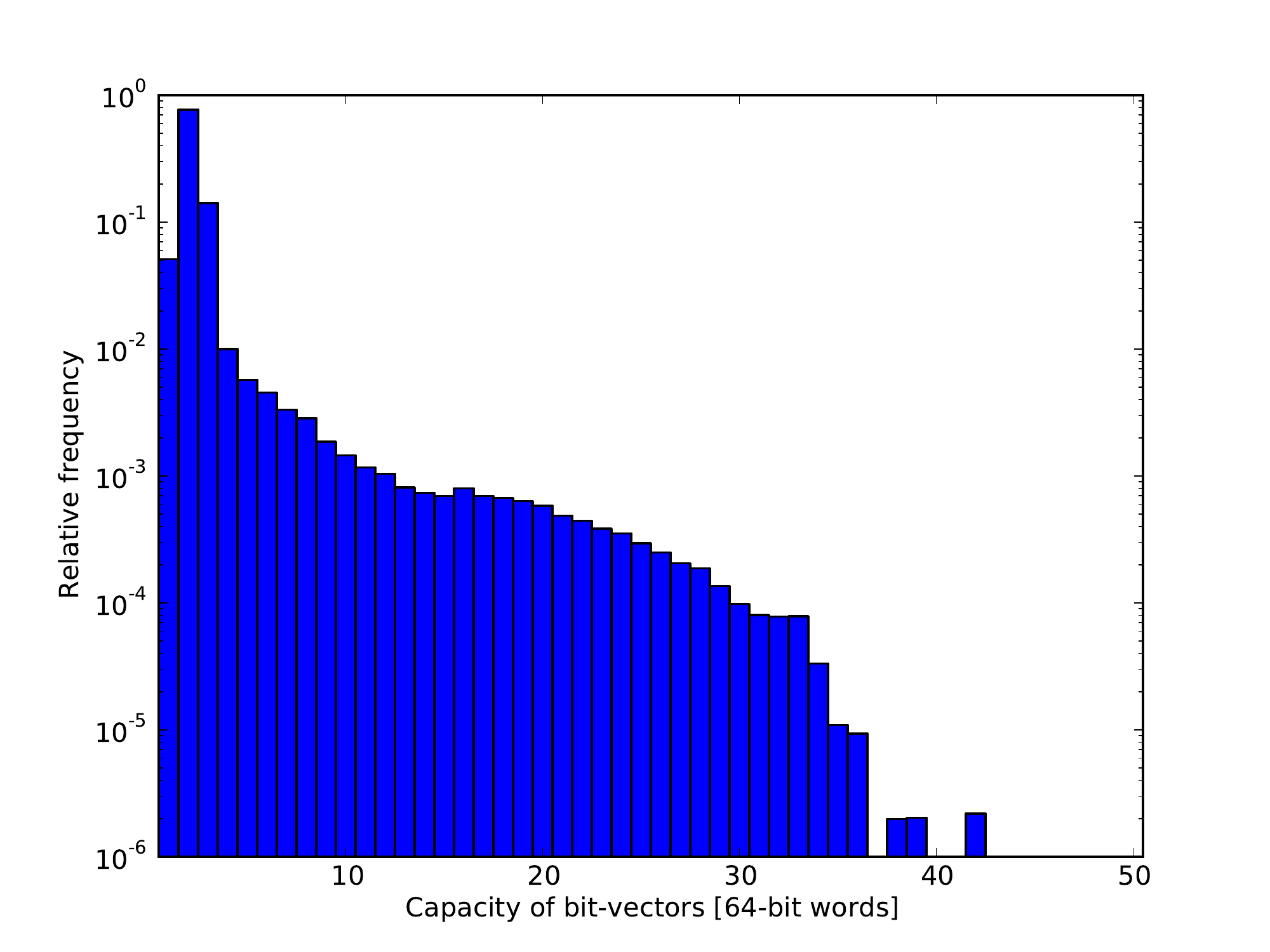}
\end{center}
\caption{Histogram of bit-vector capacities for data set NA18507.}\label{fig:NA18507_bv_capacity}
\end{figure}

The key idea leading to a practically fast algorithm is to represent
each active clique as a \emph{bitvector}, where each bit indicates
whether a particular node is part of the clique or not. We keep all
nodes from active cliques in a binary search tree sorted by their
segment length, such that vertices whose alignment satisfy condition
\eqref{eq.edgetest} can be found efficiently. We test each one of
these candidate vertices for condition \eqref{eq.edgetest2} to
identify the neighborhood of the current vertex $u$.  With our
representation of the current cliques as bitvectors, we can compute
all the intersections with $N(u)$ by a bitparallel boolean operation.

In the following we call a node \emph{active} if it is contained in at
least one active clique, otherwise we call it \emph{inactive}. When
nodes become inactive, a reorganization of all bitarrays is required
and doing this in each iteration would cancel the benefits of the fast
bitparallel operations.  To have a good trade-off between
not-too-frequent memory reorganizations and not-too-large active sets
of nodes, we employ the following strategy.  We start with a bitvector
capacity equaling the machine word size (usually 64 bits) and reserve
this amount of memory for each clique (although fewer nodes are active
in the beginning). Whenever the number of active nodes reaches the
capacity, we reorganize the data structure.  That is, we discard all
now inactive nodes, set the new capacity to twice the size of the set
of now active nodes, and repack all bitvectors.

We bound local alignment coverage by first removing alignments of
interval length $\ge 50,000$, due to that discovery of deletions of
that size is considered rather easy \cite{Alkan2011}. We further
remove alignments of weight $\Prob_0(A)< 1/625$ if necessary,
motivated by that we allow at most $25$ alignments per read end that
is $25^2=625$ alignments per paired-end read (see Results). We found
that these restrictions result in at most $\approx 500$ alignments
also in heavily repetitive areas. 

To complement the theoretical analysis in the following subsection
\ref{ssec:runtime-analysis}, we give
histograms of observed numbers of active cliques, number of nodes in
active cliques, and needed bit-vector capacities in the following.
For each run of CLEVER, these three quantities were stored in each
iteration (i.e.\ after processing another alignment pair) to obtain
the shown histograms.  Note that due to the delayed discard of
inactive nodes and the doubling of the bitvector size as described
above, active node capacities larger than 500 as can be seen in
histograms \ref{fig:venter_active_nodes} and
\ref{fig:NA18507_active_nodes}.

\begin{sloppypar}
  In our experiments, our algorithm computed 647,944,355 cliques from
  the Venter alignments and 1,193,764,528 cliques from the NA18507
  alignments.
\end{sloppypar}

\subsection{Runtime Analysis}
\label{ssec:runtime-analysis}

Please revisit paragraph 'Enumerating Maximal Cliques' in the main
text for notations and definitions in the following.

Sorting the nodes takes $\mathcal{O}(m\log m)$ time.  The intersection
of the neighborhood of the current vertex with all active cliques can
be determined by first iterating over all vertices in active cliques
and then intersecting the resulting neighborhood with each active
clique by iterating over all vertices contained in the clique.  If we
let $k$ be an upper bound on the local alignment coverage and $c$ be
the maximum number of active cliques, computing the intersection of
the neighborhood with all active cliques takes time $\mathcal{O}(kc)$.
Similarly, to detect duplicates and cliques that are subsets of other
cliques, we compute the intersection between all pairs of cliques
modified according to rule (i) or added following rule (ii).  Only
among those cliques duplicates and subcliques can arise.  Bounding the
set of new cliques by $c$ and by applying the same argument as above,
all pairwise intersections can be computed in time
$\mathcal{O}(kc^2)$.  This gives a total running time of
$\mathcal{O}(m(\log m+kc^2)+s)$, where $s$ is the size of the output.

In our experiments we bound the local alignment coverage by $k=500$,
which we achieve almost entirely through a very weak restriction of
read alignments, see above for details.
Concerning the number of active cliques, we observe a rapid drop in
relative frequency for larger sets (Supplementary
Fig.~\ref{fig:venter_active_cliques}
and~\ref{fig:NA18507_active_cliques}).  In particular, only in 1\% of
the cases the set of active cliques is larger than 268. Furthermore,
our experiments show that the number of nodes per active cliques is
typically considerably smaller than the upper bound of 500 nodes
imposed by the bound on the local alignment coverage (Supplementary
Fig.~\ref{fig:venter_active_nodes}
and~\ref{fig:NA18507_active_nodes}). In fact, in more than 99\% of the
cases the number of nodes in active cliques is smaller than 236.
These observations are one reason for the discrepancy between the
moderate worst case guarantees of our algorithm and its excellent
practical behavior.

The other reason is a careful engineering of our algorithm that
considerably improves its practical performance, although it has no or
a rather limited effect on its worst case analysis (see
Supplement~\ref{sec:algor-engin} for details).  For example, the read
alignment graph is never constructed explicitly.  Rather, we compute
the edges on demand.  Since the computation of intersections between
sets of nodes, either between the neighborhood of a new node and the
active cliques, or between pairs of new active cliques, is the key
step of our algorithm, we use a binary search tree in combination with
bit-parallel operations.  The latter cause the time required to
compute intersections between sets of nodes to drop by a factor $w$,
the word size used in the bit-parallel operations. In other words, the
time to compute $N(A)$ and $N(A)\cap \mathcal{C}$ for all active
cliques $\mathcal{C}$ reduces to $\mathcal{O}(k+\frac{k}{w}c)$, the
time required to compute the intersections between all pairs of new
active cliques reduces to $\mathcal{O}(\frac{k}{w}c)$, resulting in a
total running time of $\mathcal{O}(m(\log m+k+\frac{k}{w}c^2)+s)$.  In
Figures~\ref{fig:venter_bv_capacity}
and~\ref{fig:NA18507_bv_capacity}, we plot the relative frequencies of
different values of $\lceil\frac{k'}{w}\rceil$, where $k'\geq k$ takes
into account our delayed update of data structures, as detailed in
Supplement~\ref{sec:algor-engin}.  In 99\% of the cases
$\lceil\frac{k'}{w}\rceil\leq 4$.  That is, only four machine words
are used to store each clique and the intersection of two cliques can
thus be computed with four elementary operations.

\section{Appendix: Approximation of \eqref{eq.indeltest1}}
\label{app.approx}
In the following, we describe how to compute reasonable approximations
$\Prob^*(\cC)$ for $\Prob_{H_0}(\cC)$ in polynomial time. Due to that
we would like to ensure to keep false discovery rate under control
when correcting for multiple testing, $\Prob^*(\cC)$ should be an {\em
  upper bound} for $\Prob_{H_0}(\cC)$.\par
Let $\cA$ be a set of alignments and $w_{max}(\cA):=\max\{w(A)\mid
A\in\cA\}$ resp.~$w_{min}(\cA):=\min\{w(A)\mid A\in\cA\}$ be the
maximum resp.~minimum weight of an alignment $A\in\cA$.  As
approximation scheme for \eqref{eq.indeltest1}, we first determine
\begin{equation}
  w_{max}(\cC):=\max\{w(A)\mid A\in\cC\} 
\end{equation}
for the clique $\cC$ in question and further ($L$ for large, $S$ for
small weight)
\begin{align}
  \cC_L &:= \{A\in\cC\mid w(A)\ge\frac12 w_{max}(\cC)\}\quad\text{and}\\
  \cC_S &:=\{A\in\cC\mid w(A)<\frac12 w_{max}(\cC)\}.
\end{align}
Let further $\cC_k\subset\cC$ be the $k$ ``most concordant''
alignments in clique $\cC$, that is, $A\in\cC_k$ iff
\begin{equation}
  |I(A)-\mu| \le  |I(B)-\mu|
\end{equation}
for at least $|\cC|-k$ alignments $B\in\cC$.  Let
$\Prob_{H_0}(\cC_k)$ be the probability that the null
hypothesis of no variant holds true given that precisely the $k$ most
concordant alignments $\cC_k$ are correct. As in the main text, we
assume that $\cC=\{A_1,...,A_n\}$ consists of $n$ alignments and we
write $A_J,J\subset\N_n:=\{1,...,n\}$ for the event that precisely the
alignments $A_j,j\in J$ are correct. By definition of $\cC_k$, for
each $J\subset\N_n$ with cardinality $|J|=k$
\begin{equation}
  \label{eq.upper1}
  \Prob_{H_0}(A_J)\le\Prob_{H_0}(\cC_k).
\end{equation}
Let further $J\subset\N_n$ be of
cardinality $|J|=k$ such that 
\begin{equation}
|\{A_j,j\in J\}\cap \cC_L|=l 
\end{equation}
that is
$l$ alignments of the $A_J,j\in J$ are from $\cC_L$ which translates to
that they have comparatively large weight $w(A_j)$. If $0\le l\le |\cC_L|$
there are $\binom{|\cC_L|}{l}\cdot\binom{|\cC_S|}{k-l}$ such subsets $J$.
For each such subset, we compute
\begin{multline}
  \label{eq.upper2}
  \Prob(A_J)\le w_{k,l}(\cC):=
  w_{max}(\cC_L)^lw_{max}(\cC_S)^{k-l}\cdot
  %...\cdot 
  (1-w_{min}(\cC_L))^{|\cC_L|-l}(1-w_{min}(\cC_S))^{|\cC_S|-(k-l)}.
\end{multline}
We compute [$\binom{m_1}{m_2}:= 0, m_2>m_1$]
\begin{equation}
  \label{eq.upper3}
  \sum_{J\subset\N_n,|J|=k}\Prob(A_J)\Prob_{H_0}( A_J)\stackrel{\eqref{eq.upper1},\eqref{eq.upper2}}{\le}
  \Prob_{H_0}(\cC_k)\cdot
  \sum_{l=0}^{|\cC_L|}w_{k,l}\cdot\binom{|\cC_L|}{l}\cdot\binom{|\cC_S|}{k-l}
\end{equation}
which overall amounts to
\begin{equation}
  \begin{split}
    \Prob&_{H_0}(\cC) = \sum_{J\subset\N_n}\Prob(A_J)\Prob_{H_0}( A_J)\\
    &= \Prob(A_{\emptyset}) + \Prob(A_{\N_n}) + \sum_{k=1}^{n-1}\sum_{J\subset\N_n,|J|=k}\Prob(A_J)\Prob_{H_0}( A_J)\\
    &\stackrel{\eqref{eq.upper3}}{\le} \prod_{j=1}^n(1-w(A_j)+\prod_{j=1}^nw(A_j)\\ 
    &\qquad + \sum_{k=1}^{n-1}\Prob_{H_0}(\cC_k)\sum_{l=0}^{|\cC_L|}w_{k,l}\cdot\binom{|\cC_L|}{l}\cdot\binom{|\cC_S|}{k-l}
  \end{split}
\end{equation}
This upper bound can be computed in polynomial time. 

As a motivation for our approximation, note that if all alignments
$A\in\cC$ have equal weight, for example most importantly in case of
only uniquely mapped alignments, the approximation yields the exact
value.

\section{Appendix: Pipeline Details}\label{sec:pipeline}
In the following, we give details on how we ran all structural-variation discovery tools, including command-line options, versions, and interpretation of output.
We ``force'' every tool to make exact predictions. 
That is, for a deletion each tool has to predict a set of start and end coordinates and for an insertion each tool has to predict a breakpoint position and a length.
If such ``exact'' calls are not provided by the specific tool, we compute them from the output as detailed below.

\subsection{SamTools}
When speaking of SamTools below, we refer to version 0.1.16 (r963:234) downloaded from \url{http://samtools.sourceforge.net}.

\subsection{Read Mapping with BWA}
We used BWA version 0.6.1-r104 obtained from \url{https://github.com/lh3/bwa}.
First, an index was created by running \texttt{bwa index} with parameter \texttt{-a bwtsw}.
The input files are two (gzipped) FASTQ files containing the first and second reads of all pairs, respectively.
Each of both files was aligned by calling \texttt{bwa aln} with default parameters.
The resulting \texttt{sai} files were combined by running \texttt{bwa sampe} with parameters \texttt{-n 25} and \texttt{-N 25} to allow up to 25~alignments per read end to be reported as \texttt{XA} tags.

\subsection{Running HYDRA}\label{sec:hydra}
HYDRA version 0.5.3 was obtained from \url{http://code.google.com/p/hydra-sv}.
To prepare a BAM file that only contains discordant read pairs, we ran
\begin{verbatim}
samtools view -h -F2 bwa-out.bam | xa2multi.pl | samtools view -S -b - 
   > hydra-in.bam
\end{verbatim}
where \texttt{xa2multi.pl} is distributed along with BWA and expands \texttt{XA} tags indicating multiply mapped reads into multiple lines in the resulting BAM file; i.e.\ it has one line per alignments instead of one line per read (end).
Next, we used BEDtools version 2.15.0 (obtained from \url{https://code.google.com/p/bedtools}) to create HYDRA input files as follows.
\begin{verbatim}
bamToBed -ed -i hydra-in.bam | pairDiscordants.py -i stdin -y 0 -z 0 
   > hydra-in.bedpe
dedupDiscordants.py -i hydra-in.bedpe -s 3 > hydra-in.dedup.bedpe
\end{verbatim}
The scripts \texttt{pairDiscordants.py} and \texttt{dedupDiscordants.py} are part of the HYDRA package.
We then called HYDRA:
\begin{verbatim}
hydra -in hydra-in.dedup.bedpe -out hydra-out.breaks -mld <mld> 
   -mno <mno> 
\end{verbatim}
where \texttt{<mld>} was set to $10\cdot\sigma$ and \texttt{<mno>} was set to $20\cdot\sigma$ and $\sigma$ is the fragment size standard deviation as (robustly) estimated by BWA.
The resulting file \texttt{hydra-out.breaks.final} was then used to extract predictions as follows.
We only retain lines where both breakpoints lie on the same chromosome (i.e.\ $\texttt{field1}=\texttt{field4}$) and breakpoint orientations are not equal (i.e.\ $\texttt{field9}\neq\texttt{field10}$).
To decide whether a prediction corresponds to an insertion or to a deletion, we compare the difference of the start of second breakpoint (\texttt{field5}) and the end of first breakpoint (\texttt{field3}) to the mean internal segment size $\mu$ as estimated by BWA.
If the difference is larger than $\mu$, we interpret the prediction as a deletion from end of first breakpoint (\texttt{field3}) to the start of second breakpoint (\texttt{field5}).
If it is smaller than $\mu$, we interpret it as an insertion at position $(\texttt{field3}+\texttt{field5})/2$ of length $\mu-(\texttt{field5}-\texttt{field3})$.

\subsection{Running GASV}

\begin{sloppypar}
We used GASV version 1.5.1 as downloaded from
\url{http://code.google.com/p/gasv}.  The script
\texttt{BAM\_preprocessor.pl} coming with GASV was run on the BAM file
produced by BWA, resulting in files named
\texttt{dataset\_all.deletion}, \texttt{dataset\_all.divergent},
\texttt{dataset\_all.inversion}, \texttt{dataset\_all.translocation},
and \texttt{dataset.info}.  Then, the main GASV program was called:
\begin{verbatim}
gasv --cluster --lmin <Lmin> --lmax <Lmax> --minClusterSize 2 
   dataset_all.deletion
\end{verbatim}
where the values \texttt{<Lmin>} and \texttt{<Lmax>} were taken from \texttt{dataset.info}.
GASV's predictions were read from the produced file \texttt{dataset\_all.deletion.clusters}.
Only lines with breakpoints on the same chromosome were used (i.e.\ $\texttt{field2}=\texttt{field4}$).
GASV does not predict insertions.
To obtain deletion calls, we took the arithmetic mean for each of the two values given in each of the two fields 3 and 5 and used these two means as start and end positions of a called deletion.
\end{sloppypar}

\subsection{Running PINDEL}
We used PINDEL version 0.2.4d obtained from \url{https://trac.nbic.nl/pindel}.
As PINDEL requires sorted BAM files and the corresponding index in a \texttt{bai} file, these were produced using \texttt{samtools sort} and \texttt{samtools index}, respectively.
PINDEL was then run using the following command line.
\begin{verbatim}
pindel -T 8 -f hg18.fasta -i pindel.config -c ALL -o pindel-out
\end{verbatim}
\begin{sloppypar}
where \texttt{pindel.config} contains the name of the sorted BAM file
and the mean fragment length as estimated by BWA.  Of the resulting
files, we interpret only \texttt{pindel-out\_D},
\texttt{pindel-out\_SI}, and \texttt{pindel-out\_LI}.  We only extract
events of length larger or equal to ten (i.e.\ $\texttt{field3}\geq
10$).  For deletions, we read the chromosome, start, and end
coordinate from fields 8, 10, and 11, respectively.  For insertions,
we read the chromosome, breakpoint positions and length from fields 8,
10, and 3, respectively.
\end{sloppypar}

\subsection{Running Breakdancer}
\begin{sloppypar}
We obtained BreakDancer version 1.1\_2011\_02\_21 from
\url{http://sourceforge.net/projects/breakdancer}.  Like PINDEL,
BreakDancer runs on sorted BAM files.  First, we ran
\texttt{bam2cfg.pl} to create a config file from the input BAM file.
Then, \texttt{breakdancer\_max} was invoked on the config file without
further parameters.  We interpreted the resulting output file as
follows.  Again, we discarded all predictions with two breakpoints
corresponding to different chromosomes
(i.e.\ $\texttt{field1}\neq\texttt{field4}$).  For all lines
indicating a deletion (i.e.\ $\texttt{field7}=\texttt{DEL}$), we
extracted chromosome, start, and end coordinate from fields 1, 2, and
5, respectively.  For all lines indicating an insertion
(i.e.\ $\texttt{field7}=\texttt{INS}$), we determined chromosome,
breakpoint position, and length as \texttt{field1},
$(\texttt{field2}+\texttt{field5})/2$, and the negative of
\texttt{field8}, respectively.
\end{sloppypar}

\subsection{Running VariationHunter}

VariationHunter (VH) version 0.3 was downloaded from
\url{http://compbio.cs.sfu.ca/strvar.htm}.  We adapted the source code
to allow parameters to be given at the command line.  These changes do
not influence the results in any way.  VariationHunter expects a
special input format called \texttt{divet}.  We ran the BAM file
produced by BWA through \texttt{xa2multi.pl} (see
Section~\ref{sec:hydra} above) and used a custom script
\texttt{bam2divet.py} to produce VH input.
\begin{verbatim}
samtools view -h bwa-out.bam | xa2multi.pl | samtools view -S -b 
   > input.bam
bam2divet.py input.bam <min> <max> > vh-in.divet
\end{verbatim}
where \texttt{<min>} and \texttt{<max>} are the parameters defining
the discordant reads, i.e. reads with internal length smaller (higher)
than min (max). We set to $\mu-4\cdot\sigma$ and $\mu+4\cdot\sigma$,
respectively, following the study describing VariationHunter.

%\todo{Ivan, could you describe what bam2divet.py does here?}

%As parameters to VH, we set the the minimum and maximum paired-end
% insert size to $\mu-4\cdot\sigma$ and $\mu+4\cdot\sigma$,
% respectively.

The ``pre-processing mapping prune probability'' is set to $1.0$ and
the minimum support for a cluster to $2$.  VH writes insertions,
deletions, and inversions to separate files.  We only consider
insertions and deletions.  For deletions, we use the fields
\texttt{Inside\_End} and \texttt{OutSide\_Start} as start and end of
the predicted deletion, respectively.  For insertions, we set the
breakpoint position to
$(\texttt{OutSide\_Start}-\texttt{Inside\_End})/2$ and the length to
$\mu-\texttt{Avg\_Span}$.

\subsection{MODIL}

Modil version 1.1 was obtained from
\url{http://compbio.cs.toronto.edu/modil/src/modil_beta_v2.tar.gz}.

Modil expects a special input format. We used a custom script to
produce MODIL input from a sorted BAM file.

\begin{verbatim}
samtools view -h bwa-out.bam | xa2multi.pl | samtools view -S -b 
   > input.bam
samtools sort input.bam input
sortedbam2modil.py input.sorted.bam
\end{verbatim}

The script generates standard output files named by chromosome, which
should be placed as indicated in the \texttt{README.txt}.

\begin{sloppypar}
  We set parameters \texttt{MEAN\_INSERT\_SIZE} and \texttt{STD\_INSERT\_SIZE} in
  the MODIL input file \texttt{mrfstructvar.properties} to $\mu$ and $\sigma$, respectively.
  Again, we use the values estimated by BWA.
  All other parameters used default values. Note that many parameters
  defines system specific and data location paths as described in
  \texttt{README.txt} from MODIL.
\end{sloppypar}

Because of the high computational requirements, we ran MODIL on a grid engine by using a 
step size of 50.000 bases at \texttt{MoDIL\_simple.py}.

\subsection{SV-seq2}
SV-seq version 2 was downloaded from \url{http://www.engr.uconn.edu/~jiz08001/svseq2.html}.
We converted the FASTA file of the reference genome such that all nucleotides where in upper case and all characters not in $\{\texttt{A},\texttt{C},\texttt{G},\texttt{T},\texttt{N}\}$ were replaced by \texttt{N}.
Again, we used the sorted BAM files as output by BWA as well as $\mu$ and $\sigma$ as determined by BWA.
SV-seq was invoked two times per chromosome, once for deletions and once for insertions, as follows.
\begin{verbatim}
SVseq2 -r reference.fasta -b bwa-out.bam -c <chromosome> 
    --is <mu> <sigma> --c 1 --o deletions-out.txt
SVseq2 -insertion -r reference.fasta -b bwa-out.bam -c <chromosome>
    --is <mu> <sigma> --c 1 --o insertions-out.txt
\end{verbatim}
From the output file \texttt{deletions-out.txt}, we extracted all lines starting with \texttt{range} and used fields 3 and 5 as start and end position of a predicted deletion.
From the output file \texttt{insertions-out.txt}, we read all lines following a line consisting solely of hashes (\texttt{\#}) and interpreted field 2 as the position of a predicted insertion.
Note that SV-seq does not predict the length of insertions.

\subsection{CLEVER}
The results reported in the paper correspond to using CLEVER 1.1 (available from \url{http://code.google.com/p/clever-sv}) and calling
\begin{verbatim}
clever-all-in-one -B bwa-out.bam hg18.fasta <result-dir>
\end{verbatim}
\begin{sloppypar}
  where the switch \texttt{-B} indicates that BWA alignments are used
  and thus \texttt{xa2multi.pl} must be run by
  \texttt{clever-all-in-one}.  The CLEVER documentation contains more
  details on what \texttt{clever-all-in-one} does and how a customized
  pipeline can be build.
\end{sloppypar}

\pagebreak

\section{Appendix: Fixed-Distance Hits}\label{sec:fixed-distance-stats}
In the main text (Table~1), a predicted deletion was counted as a
\emph{hit} when it overlapped a true deletion and its length deviated
by at most 100\,bp.  For insertions, we applied an analogous
criterion: a predicted insertion was counted as a hit when the
breakpoint position did not deviate by more than its length and when
its length did not deviate by more than 100\,bp. While overlap has
been a most predominant criterion in the literature, there are other
reasonable hit criteria, motivated by that predicted breakpoints can
be far apart from true breakpoints for long indels, despite that they
overlap. We therefore offer alternative statistics in the following.

In this section, we define a \emph{hit} in terms of the absolute
distance of the breakpoints of a true and a predicted insertion.  For
deletions, we use the absolute distance of center points of truth and
prediction.  For long deletions whose length is larger than the
distance threshold, this criterion is stricter than just requiring at
least one base pair overlap as before.  For short deletions, this
criterion can be more relaxed as not overlapping but close hits might
now be counted.  As before, we still require the length deviation to
be below a given threshold.  In the following, we give two tables, one
with \emph{high-accuracy calls} and one with \emph{relaxed calls}
based on using threshold 20 and 100, respectively.  In each case, the
same threshold is used for allowed distance and allowed length
deviation.

Before showing results, we make some considerations regarding
statistical significance of variant calls meeting these new criteria.
For Venter's genome, there is a total number of $32\,737$ insertions
and $31\,904$ deletions of length at least 10.  All shorter variants
were ignored in our evaluation.  For threshold 20, there are thus at
most $41\cdot 32\,737$ positions that qualify as correctly predicted
insertion breakpoints.  Guessing a breakpoint uniformly at random
would thus be successful with a probability of at most $4.4\cdot
10^{-4}$.  For insertions, this value computes to $4.2\cdot 10^{-4}$.
Even for threshold 100, we obtain random success probabilities around
$0.002$ for both insertions and deletions.  That means that even the
indel calls made with the relaxed threshold of $100$ are statistically
significant.  Note that this consideration does not take the length of
the predictions into account.  The probability of making a random
prediction with valid distance \emph{and} valid length difference is
even smaller.

\subsection{High-Accuracy Calls (Threshold 20)}
\subsubsection{Venter}
\begin{center}
\begin{tabularx}{\textwidth}{lRRRRp{10pt}RRRR}
\hline\\[-.7em]
\textbf{Method}    & \multicolumn{4}{c}{\hspace{-1em}\textbf{Venter Insertions}} & & \multicolumn{4}{c}{\hspace{-1em}\textbf{Venter Deletions}} \\
${}^*$ \footnotesize{Split-read aligner} & Prec. & Rec. & Exc. & F & & Prec. & Rec. & Exc. & F \\[.3em]
\hline\\[-.7em]
\multicolumn{10}{l}{\textbf{Length Range 20--49} (8,786 true ins., 8,502 true del.)}\\[.3em]
CLEVER          &   \textit{39.6}&  \textbf{43.0} &  \textbf{24.5} &  \textit{41.2} & &  \textit{40.2} &  \textbf{59.6} &  \textbf{21.6} &  \textbf{48.0} \\[.3em]
BreakDancer     &             -- &           0.2  &           0.0  &             -- & &          18.9  &           0.9  &           0.0  &           1.8  \\
GASV            &            N/A &            N/A &            N/A &            N/A & &           4.3  &          19.8  &           2.3  &           7.0  \\
HYDRA           &           0.0  &           0.0  &           0.0  &             -- & &             -- &           0.0  &           0.0  &             -- \\
VariationHunter &           4.9  &           0.7  &           0.1  &           1.2  & &          16.1  &           2.2  &           0.3  &           3.9  \\[.3em]
PINDEL${}^*$    &  \textbf{53.3} &          38.9  &          20.6  &  \textbf{45.0} & &          41.9  &          45.5  &          13.2  &          43.7  \\
SV-seq2${}^*$   &            N/A &            N/A &            N/A &            N/A & &  \textbf{84.0} &           0.5  &           0.0  &           1.1  \\[.3em]
\hline\\[-.7em]
\multicolumn{10}{l}{\textbf{Length Range 50--99} (2,024 true ins., 1,822 true del.)}\\[.3em]
CLEVER          &          19.7  &  \textbf{73.4} &  \textbf{22.3} &  \textbf{31.1} & &          26.4  &  \textbf{67.8} &  \textbf{30.8} &  \textbf{38.0} \\[.3em]
BreakDancer     &  \textit{33.0} &          25.9  &           0.6  &          29.0  & &  \textit{29.4} &          14.5  &           0.0  &          19.5  \\
GASV            &            N/A &            N/A &            N/A &            N/A & &          18.2  &          15.6  &           1.0  &          16.8  \\
HYDRA           &           0.0  &           0.0  &           0.0  &             -- & &             -- &           0.1  &           0.0  &             -- \\
VariationHunter &          11.1  &          42.0  &           2.7  &          17.5  & &          18.7  &          13.1  &           0.6  &          15.4  \\[.3em]
PINDEL${}^*$    &  \textbf{42.4} &           9.0  &           0.4  &          14.8  & &          40.8  &          25.9  &           1.0  &          31.7  \\
SV-seq2${}^*$   &            N/A &            N/A &            N/A &            N/A & &  \textbf{51.9} &          13.5  &           0.3  &          21.4  \\[.3em]
\hline\\[-.7em]
\multicolumn{10}{l}{\textbf{Length Range 100--50\,000} (3,101 true ins., 2,996 true del.)}\\[.3em]
CLEVER          &          19.3  &   \textbf{6.5} &   \textbf{2.6} &   \textbf{9.7} & &  \textit{47.7} &  \textbf{52.4} &   \textbf{8.4} &  \textbf{49.9} \\[.3em]
BreakDancer     &          14.1  &           4.8  &           2.2  &           7.1  & &          30.8  &          32.7  &           0.1  &          31.7  \\
GASV            &            N/A &            N/A &            N/A &            N/A & &           0.6  &          32.5  &           0.3  &           1.1  \\
HYDRA           &           0.0  &           0.0  &           0.0  &             -- & &          34.6  &          29.5  &           0.1  &          31.9  \\
VariationHunter &  \textbf{21.6} &           3.1  &           0.4  &           5.4  & &          24.2  &          29.1  &           0.5  &          26.4  \\[.3em]
PINDEL${}^*$    &             -- &           0.0  &           0.0  &             -- & &  \textbf{76.1} &          36.8  &           0.4  &          49.6  \\
SV-seq2${}^*$   &            N/A &            N/A &            N/A &            N/A & &          64.3  &          30.9  &           0.7  &          41.7  \\[.3em]
\hline
\end{tabularx}
\end{center}

\pagebreak

\subsubsection{NA18507}
\begin{center}
\begin{tabularx}{0.9\textwidth}{lRRRp{10pt}RRR}
\hline\\[-.7em]
\textbf{Method}    & \multicolumn{3}{c}{\textbf{NA18507 Insertions}} & & \multicolumn{3}{c}{\textbf{NA18507 Deletions}} \\
${}^*$ \footnotesize{Split-read aligner} & RPr. & Rec. & Exc. & & RPr. & Rec. & Exc. \\[.3em]
\hline\\[-.7em]
\multicolumn{8}{l}{\textbf{Length Range 20--49} (2,295 true ins., 2,192 true del.)}\\[.3em]
CLEVER          &   \textit{5.7} &  \textit{19.5} &   \textit{7.9} & &   \textit{5.9} &  \textit{39.7} &  \textit{9.0}  \\[.3em]
BreakDancer     &             -- &           0.0  &           0.0  & &           2.1  &           0.7  &           0.1  \\
GASV            &            N/A &            N/A &            N/A & &           0.7  &          11.2  &           1.5  \\
HYDRA           &           0.0  &           0.0  &           0.0  & &             -- &           0.0  &           0.0  \\
VariationHunter &           0.2  &           0.4  &           0.1  & &           1.6  &           1.2  &           0.1  \\[.3em]
PINDEL${}^*$    &  \textbf{12.0} &  \textbf{37.4} &  \textbf{25.9} & &           8.8  &  \textbf{61.3} &  \textbf{29.1} \\
SV-seq2${}^*$   &            N/A &            N/A &            N/A & &  \textbf{12.1} &           0.6  &           0.1  \\[.3em]
\hline\\[-.7em]
\multicolumn{8}{l}{\textbf{Length Range 50--99} (303 true ins., 294 true del.)}\\[.3em]
CLEVER          &           0.7  &  \textbf{55.8} &  \textbf{18.5} & &           1.9  &  \textbf{68.7} &  \textbf{28.6} \\[.3em]
BreakDancer     &   \textit{1.9} &           3.3  &           0.3  & &   \textit{3.4} &          14.3  &           0.0  \\
GASV            &            N/A &            N/A &            N/A & &           0.3  &           6.8  &           0.3  \\
HYDRA           &           0.0  &           0.0  &           0.0  & &             -- &           0.0  &           0.0  \\
VariationHunter &           0.3  &          35.0  &           3.0  & &           1.5  &          20.1  &           0.0  \\[.3em]
PINDEL${}^*$    &   \textbf{7.3} &          17.8  &           2.0  & &           4.7  &          34.0  &           0.0  \\
SV-seq2${}^*$   &            N/A &            N/A &            N/A & &   \textbf{6.3} &          22.8  &           0.3  \\[.3em]
\hline\\[-.7em]
\multicolumn{8}{l}{\textbf{Length Range 100--50\,000} (165 true ins., 414 true del.)}\\[.3em]
CLEVER          &           0.1  &           7.3  &           3.0  & &   \textit{3.3} &  \textbf{54.1} &   \textbf{3.6} \\[.3em]
BreakDancer     &           0.1  &           3.6  &           1.2  & &           2.5  &          38.6  &           0.2  \\
GASV            &            N/A &            N/A &            N/A & &           0.1  &          42.3  &           0.0  \\
HYDRA           &           0.0  &           0.0  &           0.0  & &           0.8  &          33.1  &           0.0  \\
VariationHunter &   \textbf{0.2} &   \textbf{7.9} &   \textbf{4.2} & &           1.2  &          32.9  &           0.5  \\[.3em]
PINDEL${}^*$    &             -- &           0.0  &           0.0  & &   \textbf{5.6} &          49.8  &           1.0  \\
SV-seq2${}^*$   &            N/A &            N/A &            N/A & &           3.5  &          31.4  &           0.2  \\[.3em]
\hline
\end{tabularx}
\end{center}

\pagebreak

\subsection{Relaxed Calls (Threshold 100)}
\label{ssec:relaxed}

\subsubsection{Venter}
\begin{center}
\begin{tabularx}{\textwidth}{lRRRRp{10pt}RRRR}
\hline\\[-.7em]
\textbf{Method}    & \multicolumn{4}{c}{\hspace{-1em}\textbf{Venter Insertions}} & & \multicolumn{4}{c}{\hspace{-1em}\textbf{Venter Deletions}} \\
${}^*$ \footnotesize{Split-read aligner}  & Prec. & Rec. & Exc. & F & & Prec. & Rec. & Exc. & F \\[.3em]
\hline\\[-.7em]
\multicolumn{10}{l}{\textbf{Length Range 20--49} (8,786 true ins., 8,502 true del.)}\\[.3em]
CLEVER          &  \textit{94.3} &  \textit{58.9} &  \textit{14.3} &  \textit{72.5} & &  \textit{93.2} &  \textbf{74.2} &   \textit{6.4} &  \textbf{82.6} \\[.3em]
BreakDancer     &             -- &           6.1  &           0.0  &             -- & &          91.0  &           8.6  &           0.0  &          15.6  \\
GASV            &            N/A &            N/A &            N/A &            N/A & &          10.3  &          45.8  &           1.3  &          16.8  \\
HYDRA           &           0.0  &           0.0  &           0.0  &             -- & &             -- &           0.1  &           0.0  &             -- \\
VariationHunter &          61.5  &          11.6  &           0.2  &          19.5  & &          82.9  &          11.3  &           0.3  &          20.0  \\[.3em]
PINDEL${}^*$    &  \textbf{95.5} &  \textbf{63.2} &  \textbf{25.3} &  \textbf{76.1} & &          68.4  &          73.9  &   \textbf{9.7} &          71.1  \\
SV-seq2${}^*$   &            N/A &            N/A &            N/A &            N/A & & \textbf{100.0} &           1.3  &           0.0  &           2.6  \\[.3em]
\hline\\[-.7em]
\multicolumn{10}{l}{\textbf{Length Range 50--99} (2,024 true ins., 1,822 true del.)}\\[.3em]
CLEVER          &          72.7  &  \textbf{88.6} &   \textbf{6.1} &  \textbf{79.9} & &          84.9  &  \textbf{82.4} &   \textbf{2.3} &  \textbf{83.6} \\[.3em]
BreakDancer     &  \textbf{90.9} &          58.5  &           0.1  &          71.1  & &  \textbf{94.6} &          52.6  &           0.1  &          67.6  \\
GASV            &            N/A &            N/A &            N/A &            N/A & &          51.5  &          39.3  &           2.2  &          44.6  \\
HYDRA           &           0.0  &           0.0  &           0.0  &             -- & &             -- &           5.1  &           0.0  &             -- \\
VariationHunter &          63.1  &          80.8  &           1.3  &          70.9  & &          76.4  &          77.4  &           1.4  &          76.9  \\[.3em]
PINDEL${}^*$    &          87.1  &          24.6  &           0.3  &          38.3  & &          76.5  &          40.0  &           0.3  &          52.5  \\
SV-seq2${}^*$   &            N/A &            N/A &            N/A &            N/A & &          88.2  &          20.9  &           0.1  &          33.7  \\[.3em]
\hline\\[-.7em]
\multicolumn{10}{l}{\textbf{Length Range 100--50\,000} (3,101 true ins., 2,996 true del.)}\\
CLEVER          &  \textbf{65.1} &          23.7  &           2.0  &          34.8  & &  \textbf{86.0} &  \textbf{68.2} &   \textbf{3.8} &  \textbf{76.1} \\[.3em]
BreakDancer     &          58.1  &          16.4  &           2.4  &          25.5  & &          65.2  &          56.8  &           0.0  &          60.7  \\
GASV            &            N/A &            N/A &            N/A &            N/A & &           0.8  &          47.9  &           0.6  &           1.6  \\
HYDRA           &           0.0  &           0.0  &           0.0  &             -- & &          70.4  &          55.7  &           0.2  &          62.2  \\
VariationHunter &          58.6  &  \textbf{25.5} &   \textbf{3.5} &  \textbf{35.5} & &          55.9  &          63.2  &           1.2  &          59.4  \\[.3em]
PINDEL${}^*$    &             -- &           2.2  &           0.0  &             -- & &          84.1  &          39.4  &           0.1  &          53.6  \\
SV-seq2${}^*$   &            N/A &            N/A &            N/A &            N/A & &          79.3  &          36.7  &           0.3  &          50.2  \\[.3em]
\hline
\end{tabularx}
\end{center}

\pagebreak

\subsubsection{NA18507}
\begin{center}
\begin{tabularx}{0.9\textwidth}{lRRRp{10pt}RRR}
\hline\\[-.7em]
\textbf{Method}    & \multicolumn{3}{c}{\textbf{NA18507 Insertions}} & & \multicolumn{3}{c}{\textbf{NA18507 Deletions}} \\
${}^*$ \footnotesize{Split-read aligner}  & RPr. & Rec. & Exc. & & RPr. & Rec. & Exc. \\[.3em]
\hline\\[-.7em]
\multicolumn{8}{l}{\textbf{Length Range 20--49} (2,295 true ins., 2,192 true del.)}\\[.3em]
CLEVER          &  \textit{10.9} &  \textit{27.6} &           7.4  & &  \textit{13.7} &  \textit{49.6} &   \textit{4.3} \\[.3em]
BreakDancer     &             -- &           0.3  &           0.0  & &          10.7  &           6.7  &           0.0  \\
GASV            &            N/A &            N/A &            N/A & &           2.5  &          41.8  &           3.9  \\
HYDRA           &           0.0  &           0.0  &           0.0  & &             -- &           0.0  &           0.0  \\
VariationHunter &           1.9  &           6.4  &           0.5  & &           6.8  &           6.4  &           0.3  \\[.3em]
PINDEL${}^*$    &  \textbf{15.6} &  \textbf{49.1} &  \textbf{29.8} & &          10.7  &  \textbf{67.9} &  \textbf{17.7} \\
SV-seq2${}^*$   &            N/A &            N/A &            N/A & &  \textbf{15.2} &           1.8  &           0.1  \\[.3em]
\hline\\[-.7em]
\multicolumn{8}{l}{\textbf{Length Range 50--99} (303 true ins., 294 true del.)}\\[.3em]
CLEVER          &           2.3  &  \textbf{73.3} &   \textbf{6.3} & &           6.8  &  \textbf{83.0} &   \textbf{3.7} \\[.3em]
BreakDancer     &   \textit{6.4} &          16.2  &           0.0  & &  \textbf{10.8} &          52.7  &           0.0  \\
GASV            &            N/A &            N/A &            N/A & &           2.8  &          43.9  &           1.4  \\
HYDRA           &           0.0  &           0.0  &           0.0  & &             -- &           2.0  &           0.0  \\
VariationHunter &           1.8  &          66.3  &           1.7  & &           5.3  &          72.8  &           1.4  \\[.3em]
PINDEL${}^*$    &  \textbf{11.4} &          32.7  &           1.0  & &           9.4  &          46.6  &           0.7  \\
SV-seq2${}^*$   &            N/A &            N/A &            N/A & &          10.4  &          29.2  &           0.0  \\[.3em]
\hline\\[-.7em]
\multicolumn{8}{l}{\textbf{Length Range 100--50\,000} (165 true ins., 414 true del.)}\\[.3em]
CLEVER          &           0.5  &          30.9  &           1.8  & &           4.7  &          66.9  &   \textbf{2.9} \\[.3em]
BreakDancer     &           0.9  &          19.4  &           0.0  & &   \textit{5.1} &          58.9  &           0.0  \\
GASV            &            N/A &            N/A &            N/A & &           0.1  &          54.6  &           1.2  \\
HYDRA           &           0.0  &           0.0  &           0.0  & &           1.6  &          60.9  &           0.5  \\
VariationHunter &   \textbf{1.7} &  \textbf{44.9} &  \textbf{11.5} & &           2.8  &  \textbf{67.4} &           1.9  \\[.3em]
PINDEL${}^*$    &             -- &           1.2  &           0.0  & &   \textbf{5.9} &          51.2  &           0.2  \\
SV-seq2${}^*$   &            N/A &            N/A &            N/A & &           3.8  &          33.6  &           0.5  \\[.3em]
\hline
\end{tabularx}
\end{center}

\pagebreak

\section{Appendix: Statistics on Length Difference and Distance}\label{sec:accuracy-stats}
In this section, we ask how accurate the valid variant calls are for
each tool.  For each prediction counted as a true positive in Table~1
in the main text (that is true positives is defined by overlap), we
computed the distance and length difference to the true indel.  In
case of deletions, distance is measured with respect to the center
points: if $(S_0,E_0)$ and $(S_1,E_1)$ are two deletions, where
$S_0,S_1$ are start coordinates and $E_0,E_1$ are end coordinates we
define their distance by
\begin{equation}
  \label{eq.centerpoint}
  {\rm Dist} := |C_0 - C_1| \quad\text{where}\quad C_i:=\frac{E_i-S_i}{2}, i=0,1
\end{equation}
Note that one can infer the distances between both start and end
coordinates from the distance of the centerpoints and the difference
in length.  The table below shows the average distances and length
differences for all tools.  A dash indicates that the respective tool
did not make any correct prediction in that category.\par In
conclusion, split-read based approaches make most accurate
predictions.  Performance rates for insert size based approaches vary
across the different categories. CLEVER usually has good rates in
terms of length difference, but has does not achieve best values in
terms of breakpoint distance.

\subsection{Venter}
\begin{center}
\begin{tabularx}{0.8\textwidth}{lRRp{10pt}RR}
\hline\\[-.7em]
\textbf{Method}    & \multicolumn{2}{c}{\textbf{Venter Insertions}} & & \multicolumn{2}{c}{\textbf{Venter Deletions}} \\
${}^*$ \footnotesize{Split-read aligner} & Dist. & Len.Diff. & & Dist. & Len.Diff. \\[.3em]
\hline\\[-.7em]
\multicolumn{6}{l}{\textbf{Length Range 20--49} (8,786 true ins., 8,502 true del.)}\\[.3em]
CLEVER          &  \textit{16.5} &   \textit{8.6} & &          14.7  &           9.6  \\[.3em]
BreakDancer     &             -- &             -- & &          19.6  &          30.4  \\
GASV            &            N/A &            N/A & & \textit{13.3}  &   \textit{7.8} \\
HYDRA           &             -- &             -- & &             -- &             -- \\
VariationHunter &          27.1  &          28.0  & &          19.7  &          32.1  \\[.3em]
PINDEL${}^*$    &  \textbf{11.5} &   \textbf{3.2} & &          10.3  &   \textbf{4.2} \\
SV-seq2${}^*$   &            N/A &            N/A & &   \textbf{9.4} &           5.0  \\[.3em]
\hline\\[-.7em]
\multicolumn{6}{l}{\textbf{Length Range 50--99} (2,024 true ins., 1,822 true del.)}\\[.3em]
CLEVER          &          28.9  &          18.8  & &          27.0  &  \textit{13.9} \\[.3em]
BreakDancer     &  \textit{22.9} &  \textit{16.4} & &  \textit{19.2} &          28.3  \\
GASV            &            N/A &            N/A & &          23.3  &          15.1  \\
HYDRA           &             -- &             -- & &             -- &             -- \\
VariationHunter &          32.8  &          28.2  & &          22.9  &          29.6  \\[.3em]
PINDEL${}^*$    &  \textbf{19.8} &   \textbf{9.0} & &  \textbf{16.3} &          10.4  \\
SV-seq2${}^*$   &            N/A &            N/A & &          16.6  &   \textbf{7.0} \\[.3em]
\hline\\[-.7em]
\multicolumn{6}{l}{\textbf{Length Range 100--50000} (3,101 true ins., 2,996 true del.)}\\[.3em]
CLEVER          &          35.9  &  \textit{23.8} & &          24.4  &          13.0  \\[.3em]
BreakDancer     &          44.9  &          23.9  & &  \textit{23.1} &          33.7  \\
GASV            &            N/A &            N/A & &          24.8  &  \textit{12.5}  \\
HYDRA           &             -- &             -- & &          24.5  &          27.8  \\
VariationHunter &  \textbf{32.6} &  \textbf{19.7} & &          30.1  &          31.8  \\[.3em]
PINDEL${}^*$    &             -- &             -- & &  \textbf{14.1} &   \textbf{4.5} \\
SV-seq2${}^*$   &            N/A &            N/A & &          21.1  &           6.4  \\[.3em]
\hline
\end{tabularx}
\end{center}
\pagebreak

\section{Appendix: External Error Sources}\label{sec:simseq}
The quality of the read alignments greatly influences the quality of
results when predicting structural variations.  For each read
generated from Venter's genome, the true origin and thus the correct
mapping location is known.  To assess the impact of false alignments
on the reported results, we ran CLEVER on the true alignments.  That
is, we converted the position of a read from donor coordinates to
reference coordinates, created a BAM file, and used it as input to
CLEVER.  The performance using these alignments is shown in the below
table (rows labeled \emph{True}).  However, even a perfect read mapper
cannot always uniquely find the correct mapping locations because of
repetitive areas in the genome.  To have a more realistic assessment
of how the results would be for a ``perfect'' read mapper, we merged
BWA alignments and true alignments.  Whenever the true alignment of a
read was not among the set of alignments reported by BWA, we added the
true alignment.  In this way, we obtained a BAM file that contains the
true alignment for every read, but additionally contains (wrong)
alternative alignments for those reads that map to multiple locations
(rows labeled \emph{BWA+True}). Losses in performance on this dataset
can be mainly attributed to mistaken alignment quality scores which in
turn point to the quality of the Phred scores involved, another
potential source of errors, in particular when dealing with multiply
mapped reads. The numbers reported in the table were computed in the
same way as the numbers in Table~1 in the main text.

\begin{center}
\begin{tabularx}{0.8\textwidth}{lRRRp{10pt}RRR}
\hline\\[-.7em]
\textbf{Alignments}    & \multicolumn{3}{c}{\textbf{Venter Insertions}} & & \multicolumn{3}{c}{\textbf{Venter Deletions}} \\
  & Prec. & Rec. & F & & Prec. & Rec. & F \\[.3em]
\hline\\[-.7em]
\multicolumn{8}{l}{\textbf{Length Range 20--49} (8,786 true ins., 8,502 true del.)}\\[.3em]
BWA       &          62.5  &  \textbf{53.0} &          57.4  & &          60.4  &          66.8  &          63.4  \\
BWA+True  &          77.6  &          35.9  &          49.1  & &  \textbf{76.4} &          53.3  &          62.8  \\
True      &  \textbf{95.8} &          41.6  &  \textbf{58.1} & &          71.5  &  \textbf{68.6} &  \textbf{70.1} \\[.3em]
\hline\\[-.7em]
\multicolumn{8}{l}{\textbf{Length Range 50--99} (2,024 true ins., 1,822 true del.)}\\[.3em]
BWA       &          60.4  &          86.6  &          71.2  & &          72.7  &          80.7  &          76.5  \\
BWA+True  &          64.2  &          77.0  &          70.1  & &          85.0  &          75.5  &          80.0  \\
True      &  \textbf{97.7} &  \textbf{93.8} &  \textbf{95.7} & &  \textbf{96.1} &  \textbf{97.3} &  \textbf{96.7} \\[.3em]
\hline\\[-.7em]
\multicolumn{8}{l}{\textbf{Length Range 100--50\,000} (3,101 true ins., 2,996 true del.)}\\[.3em]
BWA       &          66.2  &  \textbf{23.8} &  \textbf{35.1} & &          87.6  &          69.9  &          77.7  \\
BWA+True  &          72.5  &          18.4  &          29.3  & &          90.9  &          75.7  &          82.6  \\
True      &  \textbf{97.3} &          12.4  &          22.0  & &  \textbf{95.5} &  \textbf{98.2} &  \textbf{96.9} \\[.3em]
\hline
\end{tabularx}
\end{center}

The above table confirms that the quality of alignments is indeed important for the prediction of structural variations.
Using the true alignments (row \emph{True}) yields the best results in most cases, but---remarkably---not in all cases.
We have not yet found an explanation for this phenomenon.

Interestingly, the difference in overall performance (as given by the F-measure) between  \emph{BWA} and \emph{BWA+True} is smaller than the difference between \emph{BWA+True} and \emph{True} in most cases.
This suggests that the negative influence of the fact that the genome is repetitive (and reads can thus map to multiple locations) is much stronger than the influence of the  imperfectness of the read mapper.

\pagebreak

\section{Appendix: MoDIL Results on Chromosome 1}
\label{sec:modil}

MoDIL shows extraordinarily long runtimes.
We therefore ran MoDIL only on simulated reads from chromosome 1 of Venter's genome.
To have an unbiased comparison, we show MoDIL's results together with the results of the other tools when run on this restricted data set.
Statistics were computed in the same way as for Table~1 in the main text.

\begin{center}
\begin{tabularx}{\textwidth}{lRRRRp{10pt}RRRR}
\hline\\[-.7em]
\textbf{Method}    & \multicolumn{4}{c}{\textbf{Venter Insertions (Chr. 1)}} & & \multicolumn{4}{c}{\textbf{Venter Deletions (Chr. 1)}} \\
${}^*$ \footnotesize{Split-read aligner}  & Prec. & Rec. & Exc. & F & & Prec. & Rec. & Exc. & F \\[.3em]
\hline\\[-.7em]
\multicolumn{10}{l}{\textbf{Length Range 20--49} (644 true ins., 640 true del.)}\\[.3em]
MoDIL           &          31.8  &          50.5  &   \textit{4.0} &          39.0  & &          34.0  &          50.3  &           2.3  &          40.5  \\[.3em]
CLEVER          &  \textit{64.2} &  \textbf{52.6} &   \textit{4.0} &  \textbf{57.9} & &          61.7  &  \textbf{68.1} &   \textit{7.8} &  \textbf{64.7} \\
BreakDancer     &             -- &           4.3  &           0.0  &             -- & &  \textbf{78.8} &           5.2  &           0.0  &           9.7  \\
GASV            &            N/A &            N/A &            N/A &            N/A & &           5.1  &          27.7  &           0.9  &           8.6  \\
HYDRA           &           0.0  &           0.0  &           0.0  &             -- & &             -- &           0.2  &           0.0  &             -- \\
VariationHunter &          34.2  &           7.8  &           0.0  &          12.7  & &          64.1  &           9.5  &           0.3  &          16.6  \\[.3em]
PINDEL${}^*$    &  \textbf{65.5} &          48.1  &  \textbf{15.8} &          55.5  & &          51.8  &          58.3  &  \textbf{10.5} &          54.8  \\
SV-seq2${}^*$   &            N/A &            N/A &            N/A &            N/A & &             -- &           0.0  &           0.0  &             -- \\[.3em]
\hline\\[-.7em]
\multicolumn{10}{l}{\textbf{Length Range 50--99} (153 true ins., 130 true del.)}\\[.3em]
MoDIL           &          41.9  &  \textbf{88.9} &   \textbf{5.9} &          57.0  & &          39.0  &          69.2  &           1.5  &          49.9  \\[.3em]
CLEVER          &          56.1  &          88.2  &           0.6  &          68.6  & &          72.3  &  \textbf{79.2} &   \textbf{3.8} &  \textbf{75.6} \\
BreakDancer     &  \textbf{85.5} &          57.5  &           0.0  &  \textbf{68.8} & &  \textbf{91.5} &          42.3  &           0.0  &          57.9  \\
GASV            &            N/A &            N/A &            N/A &            N/A & &          51.3  &          37.7  &           1.5  &          43.4  \\
HYDRA           &             -- &           0.0  &           0.0  &             -- & &             -- &           6.9  &           0.8  &             -- \\
VariationHunter &          50.4  &          75.2  &           0.6  &          60.3  & &          67.0  &          59.2  &           0.8  &          62.9  \\[.3em]
PINDEL${}^*$    &          83.3  &          22.9  &           0.0  &          35.9  & &          67.9  &          35.4  &           0.0  &          46.5  \\
SV-seq2${}^*$   &            N/A &            N/A &            N/A &            N/A & &             -- &           0.0  &           0.0  &             -- \\[.3em]
\hline\\[-.7em]
\multicolumn{10}{l}{\textbf{Length Range 100--50000} (223 true ins., 198 true del.)}\\[.3em]
MoDIL           &          43.1  &  \textbf{22.4} &   \textbf{4.0} &          29.5  & &          60.0  &          11.6  &           1.5  &          19.5  \\[.3em]
CLEVER          &          71.9  &          15.7  &           0.0  &          25.8  & &  \textbf{92.6} &  \textbf{70.2} &   \textbf{3.0} &  \textbf{79.9} \\
BreakDancer     &          58.6  &          12.6  &           2.2  &          20.7  & &          67.8  &          62.6  &           0.0  &          65.1  \\
GASV            &            N/A &            N/A &            N/A &            N/A & &           0.7  &          50.0  &           0.5  &           1.4  \\
HYDRA           &             -- &           0.0  &           0.0  &             -- & &          69.3  &          63.6  &           1.0  &          66.4  \\
VariationHunter &  \textbf{83.3} &          20.2  &           0.4  &  \textbf{32.5} & &          64.8  &          65.7  &           0.0  &          65.2  \\[.3em]
PINDEL${}^*$    &             -- &           1.4  &           0.0  &             -- & &          86.2  &          48.5  &           0.0  &          62.1  \\
SV-seq2${}^*$   &            N/A &            N/A &            N/A &            N/A & &             -- &           0.0  &           0.0  &             -- \\[.3em]
\hline
\end{tabularx}
\end{center}

\bibliographystyle{natbib}
\bibliography{CleverShort}

\end{document}